\documentclass[aps,prb,amsmath,amssymb,twocolumn]{revtex4}
\usepackage[utf8]{inputenc}
\usepackage{graphicx}
\usepackage{dcolumn}% Align table columns on decimal point
\usepackage{bm}% bold math

\newcommand{\Tr}{\mathop\mathrm{Tr}}
\renewcommand{\Re}{\mathop\mathrm{Re}}
\renewcommand{\Im}{\mathop\mathrm{Im}}
\newcommand{\bz}{\textsc{bz}}
\newcommand{\ir}{\textsc{ir}}

\usepackage{xcolor} 
\usepackage{hyperref}
\begin{document}

%\preprint{APS/123-QED}

\title{A random matrix approach to the boson peak\\ and Ioffe-Regel criterion in amorphous solids}
\author{D.\,A. Conyuh}
\email{conyuh.dmitrij@yandex.ru}
\author{Y.\,M. Beltukov}%
\affiliation{Ioffe Institute, 194021 Saint-Petersburg, Russia}
%\affiliation{Peter the Great St.Petersburg Polytechnic University, 195251 Saint-Petersburg, Russia}
%\collaboration{}%\noaffiliation
\date{\today}

\begin{abstract}
    We present a random matrix approach to study general vibrational properties of stable amorphous solids with translational invariance using the correlated Wishart ensemble. Within this approach, both analytical and numerical methods can be applied. Using the random matrix theory, we found the analytical form of the vibrational density of states and the dynamical structure factor.  We demonstrate the presence of the Ioffe-Regel crossover between low-frequency propagating phonons and diffusons at higher frequencies. The reduced vibrational density of states shows the boson peak, which frequency is close to the Ioffe-Regel crossover. We also present a simple numerical random matrix model with finite interaction radius, which properties rapidly converges to the analytical results with increasing the interaction radius. For fine interaction radius, the numerical model demonstrates the presence of the quasilocalized vibrations with a power-law low-frequency density of states.
\end{abstract}

\maketitle

\section{Introduction}
Establishing the general vibrational properties in amorphous dielectrics (glasses) is one of the key problems in the physics of disordered systems. The dominant part of the vibrational spectrum above the Ioffe-Regel crossover and below the mobility edge is occupied by \emph{diffusons}~\cite{Allen 1993, Allen 1999}. These delocalized vibrations are spread by means of diffusive energy transfer from atom to atom. The diffusons correspond to heat transfer in glasses in a wide range of temperatures. Therefore, it is important to understand the microscopic nature of these vibrational modes.

Another universal vibrational property of almost all glasses is an excess vibrational density of states (VDOS) well-known as the \emph{boson peak}. The boson peak was observed using different experimental techniques: the Raman scattering~\cite{Malinovsky 1986, Kabeya 2016}, the X-ray scattering~\cite{Benassi 1996}, the inelastic neutron scattering~\cite{Wischnewski 1998}, the far-infrared spectroscopy~\cite{Matsuishi 1986, Hutt 1989, Ohsaka 1994}, and the temperature dependence of the heat capacity~\cite{Zeller 1971, Phillips 1981, White 1984, Kojima 2011}. Also, the boson peak appears in two-dimensional structures~\cite{Steurer 2007, Steurer 2008-1, Steurer 2008-2, Zhang 2017, Wang 2018}. It was observed that the boson peak frequency $\omega_b$ is usually close to the frequency $\omega_\ir$ of the Ioffe-Regel crossover between well-defined phonons with a long mean free path and disordered vibrations, diffusons~\cite{Ruffle 2006, Ruffle 2008, Shintani 2008}. Therefore, a general theory of the boson peak and Ioffe-Regel criterion can shed light on the nature of vibrations in amorphous solids.

There are several theoretical explanations of these anomalies such as an effective medium theory of elasticity~\cite{Schirmacher 2007, Marruzzo 2013, DeGiuli 2014, DeGiuli 2015, Wyart 2010, Wyart 2005}, soft-potential model~\cite{Gurevich 2003, Parshin 2007, Buchenau 1991, Karpov 1983, Buchenau 1992}, the mode-coupling theory~\cite{Geotze 2000}, attribution to the transverse-acoustic van Hove singularity~\cite{Taraskin 2001, Chumakov 2011, Tanaka 2008} and the breaking of the local inversion symmetry~\cite{Milkus 2016}. Hence, it is important to find the most general properties of amorphous solids, which are relevant to the formation of the boson peak and the Ioffe-Regel criterion. 

To study these fundamental vibrational features of amorphous solids, we use an approach based on the random matrix theory (RMT). This theory has important applications in many diverse areas of science and engineering~\cite{Speicher 2012, Prahofer 2000, Laloux 2000, Rajan 2006, Harnad 2011, Tulino 2004, Wage 2015, Meyer 1997, Guhr 1998, Olekhno 2018}. Depending on the inherent symmetry properties of different disordered systems, various random matrix ensembles are used. Vibrations of amorphous solids are characterized by eigenvalues and eigenvectors of the dynamical matrix $\hat{M}$. The presence of disorder in amorphous systems leads to the random nature of the matrix elements $M_{ij}$. Therefore, the RMT can be applied to study vibrational properties of amorphous solids~\cite{Grigera 2002, Manning 2015, Beltukov 2013, Baggioli 2019}. It is also applicable to jammed solids~\cite{Beltukov 2015}, which are widely studied nowadays~\cite{Liu 2010, OHern 2003}. However, not every random matrix ensemble takes into account special correlations between matrix elements $M_{ij}$ in amorphous solids. In this work we consider a \emph{correlated} ensemble, which takes into account only two of the most important properties of amorphous solids: (i) the system is near the \emph{stable} equilibrium position and (ii) the potential energy is invariant under the continuous \emph{translation} of the system. We demonstrate that the general properties (i) and (ii) determine a correlated random-matrix ensemble, which represents the most important properties of amorphous solids like the boson peak and the Ioffe-Regel crossover. 

The paper is organized as follows. In Section~\ref{sec:Wishart}, we present the correlated Wishart ensemble of the form $\hat{M} = \hat{A}\hat{A}^T$ with the sum rule $\sum_i A_{ik} = 0$, which is the most general ensemble, which satisfies the symmetrical properties (i) and (ii). Using the random matrix theory, we find the general spectral properties of the dynamical matrix $\hat{M}$. In Section~\ref{sec:num}, we present a numerical random matrix model, which refines the given correlated Wishart ensemble for a finite interaction radius. Sections~\ref{sec:VDOS} and~\ref{sec:DSF} demonstrate the vibrational density of states and the dynamical structure factor obtained from the general spectral properties of the dynamical matrix $\hat{M}$. Section~\ref{sec:IRC} demonstrates the presence of the Ioffe-Regel crossover, which splits the vibrational spectrum into phonon and diffuson frequency ranges with different vibrational properties. In Section~\ref{sec:QLV}, we demonstrate the presence of the quasilocalized vibrations with the low-frequency power-law density of states, which is determined by the non-Gaussian properties of the random matrix $\hat{A}$. Finally, in Section~\ref{sec:Disc} we discuss the obtained results and compare them with experiments and other theories.

\section{Correlated Wishart ensemble}\label{sec:Wishart}

The mechanical stability of amorphous solids is equivalent to the positive definiteness of the dynamical matrix $\hat{M}$. Any positive definite matrix $\hat{M}$ can be written as $\hat{M} = \hat{A}\hat{A}^T$ and vice versa, $\hat{A}\hat{A}^T$ is positive definite for any (not necessarily square) matrix $\hat{A}$~\cite{Bhatia}. Therefore, we can consider a $N\times K$ random matrix $\hat{A}$ to obtain a mechanically stable system with the dynamical matrix in the form of the Wishart ensemble $\hat{M}=\hat{A}\hat{A}^T$. Each column of the matrix $\hat{A}$ represents a \emph{bond} with non-negative potential energy~\cite{Beltukov 2016}
\begin{equation}
    U_k = \frac{1}{2}\Big(\sum_{i}A_{ik}u_i\Big)^2,   \label{eq:bond}
\end{equation}
where $u_i$ is a displacement of $i$-th degree of freedom from the equilibrium value. Usually, only several elements $A_{ik}$ are nonzero in each column of the matrix $\hat{A}$. In the simplest case there are only two nonzero elements, which represent a simple spring with pairwise interaction with an energy $U_{ij} = \frac{k_{ij}}{2}(u_i - u_j)^2$. The bond defined by Eq.~(\ref{eq:bond}) is a generalization of the simple spring to the many-body interaction. Each bond with many-body interaction can be represented as a number of simple springs. However, a number of them may have negative stiffness $k_{ij}$, which complicates further analysis due to the individual instability of such springs~\cite{Beltukov 2013}. 

For simplicity, we consider a scalar model with unit masses $m_i=1$. In this case the displacements of atoms are described by scalars $u_i$. The bond energy $U_k$ should not depend on the continuous translation $u_i \to u_i + const$. Therefore, the matrix $\hat{A}$ obeys the \emph{sum rule} $\sum_i A_{ik} = 0$. In the framework of the random matrix theory, it means that the matrix elements $A_{ij}$ are \emph{correlated}. 

In the minimal model, we can assume that amorphous solid consists of statistically equivalent random bonds. In this case the pairwise correlations between matrix elements $A_{ij}$ can be written as
\begin{equation}
    \langle A_{ik}A_{jl} \rangle = \frac{1}{N}C_{ij}\delta_{kl},  \label{eq:cor}
\end{equation}
where $\hat{C}$ is some correlation matrix. One can see that the correlation matrix $\hat{C}$ is proportional to the average dynamical matrix: $\hat{C} = \tfrac{N}{K}\langle\hat{M}\rangle$. For simplicity, we consider a model amorphous solid as a simple cubic lattice with random bonds and unit lattice constant $a_0=1$. In this case the \emph{average} dynamical matrix $\langle\hat{M}\rangle$ describes the simple cubic crystal. It is natural to assume that the crystalline matrix $\langle\hat{M}\rangle$ has simple bonds between the nearest neighbors with a certain rigidity. In this case the matrix $\hat{C}$ has the following structure. Non-diagonal elements are $C_{ij}=-\Omega^2$ if atoms with indices $i$ and $j$ are the nearest neighbors in the lattice and $C_{ij} = 0$ otherwise. Diagonal elements are $C_{ii}=6\Omega^2$. The constant $\Omega$ defines a typical frequency in the system. Thus, the correlation matrix $\hat{C}$ is a regular matrix, which describes a simple cubic lattice with the nearest neighbor interaction. Eigenvalues of the matrix $\hat{C}$ depend on the wavevector ${\bf q}$ which can be expressed as a dispersion law
\begin{equation}
    \omega_0^2({\bf q})=4\Omega^2\Big(\sin^2\frac{q_x}{2} + \sin^2\frac{q_y}{2} + \sin^2\frac{q_z}{2}\Big).   \label{eq:disp}
\end{equation}

Using the random matrix approach, it can be shown that statistical properties of the random matrix $\hat{M}$ are related to the known correlation matrix $\hat{C}$. To find these properties, we consider the corresponding resolvents:
\begin{equation}\label{eq:Res}
    \hat{G}(z) = \left<\frac{1}{z - \hat{M}}\right>, \quad
    \hat{G}_0(Z) = \frac{1}{Z - \hat{C}},
\end{equation}
where $z$ and $Z$ are complex parameters. The averaging is performed over different realizations of the random matrix $\hat{M}$. In the thermodynamic limit $N\to\infty$ there is a fundamental duality relation between spectral properties of $\hat{M}$ and $\hat{C}$~\cite{Burda 2004}:
\begin{equation}
    Z\hat{G}_0(Z) = z\hat{G}(z),   \label{eq:GG}
\end{equation}
where complex parameters $z$ and $Z$ are related by a conformal map $Z(z)$ defined by the equation
\begin{equation}
    \varkappa Z + \frac{Z^2}{N}\Tr \hat{G}_0(Z) = z.   \label{eq:conformal}
\end{equation}
Parameter $\varkappa = (K-N)/N$ defines the relative excess of the total number of bonds in comparison to the total number of degrees of freedom. The duality relation (\ref{eq:GG}) makes it possible to find the vibrational density of states (VDOS) and the dynamical structure factor (DSF) of the dynamical matrix $\hat{M}$.

The difference between the total number of bonds $K$ and the total number of degrees of freedom $N$ plays a crucial role in vibrational and mechanical properties. In a stable system with a finite rigidity, the number of bonds should be larger than the number of degrees of freedom, which is known as the Maxwell counting rule~\cite{Maxwell 1865}. Therefore, the parameter $\varkappa$ varies in a wide range $0 \leq \varkappa < \infty$ and controls the relationship between stiffness and disorder in the system. 

For the jammed solids, it was shown that many properties (like the shear modulus and the crossover frequency) are scaled with $\varkappa$~\cite{OHern 2003, Beltukov 2015}. In this model, the number of bonds $K$ is defined by the total number of contacts between granules. Using the variation of the density of jammed packing, the value of the parameter $\varkappa$ can be varied in a wide range of values $0 < \varkappa \lesssim 1$.

In amorphous solids, the parameter $\varkappa$ could not be varied in such a wide range. However, different amorphous solids have different typical values of the parameter $\varkappa$. In the Stillinger-Weber model of amorphous silicon, each atom has $4/2=2$ bonds determining the distance between atoms and 6 bonds determining the angles between the bonds of amorphous silicon~\cite{Stillinger 1985}. Therefore, we can estimate the parameter as $\varkappa = 5/3 \sim 1$. In amorphous SiO$_2$, there are 9 degrees of freedom per each silicon atom. In this case, the number of bonds can be estimated as 12 (4 determine the Si-O distance, 6 determine the O-Si-O angles, and $4/2 = 2$ determine the Si-O-Si angles). In this case $\varkappa=1/3$. Actually, the Si-O-Si angle bond has a lower stiffness, which results in a slightly smaller typical value of the parameter $\varkappa$. The difference in parameters $\varkappa$ for amorphous silicon and SiO$_2$ correlates well with the difference of vibrational density of states, the boson peak frequency, and the Ioffe-Regel criterion in these systems. 

The derivation of the duality relation (\ref{eq:GG}) assumes that the matrix $\hat{A}$ has the multivariate Gaussian distribution~\cite{Burda 2004}
\begin{equation}
    p(\hat{A}) \sim \exp\left(-\frac{1}{2}\Tr \hat{A}^T \hat{C}^{-1}\hat{A}\right),
\end{equation}
which implies the long-range interaction between atoms. However, the result (\ref{eq:GG}) may be used for a much wider class of sparse random matrices $\hat{A}$, which corresponds to a short-range interaction between atoms. Before investigating the vibrational properties using the duality relation (\ref{eq:GG}), we present a numerical model with short-range interaction and the same correlation matrix $\hat{C}$.

\section{The numerical model}
\label{sec:num}

In this section we present the numerical model of an amorphous system as a lattice with random bonds between atoms.

For $\varkappa=0$ the matrix $\hat{A}$ is square and the number of bonds $K$ is equal to the number of degrees of freedom $N$. We assume that the bonds cover the lattice uniformly, which results in one bond per atom in the scalar model under consideration. Thus, we can consider the following structure of the non-diagonal elements of the matrix $\hat{A}$:
\begin{equation}
    A_{ij} = \left\{
    \begin{array}{ll}
        \frac{\Omega}{2}\xi_{ij} & \text{if $i$ and $j$ are neighbors}, \\
        0 & \text{otherwise},
    \end{array}
    \right.  \label{eq:A_num}
\end{equation}
where $\xi_{ij}$ are independent Gaussian random numbers with zero mean and unit variance. The diagonal elements are defined using the sum rule $A_{ii} = -\sum_{j\neq i} A_{ji}$. Up to the normalization constant, this structure of the square random matrix $\hat{A}$ was considered in~\cite{Beltukov 2013}. The relation between the matrix $\hat{A}$ and the dynamical matrix $\hat{M}$ is also discussed in detail in~\cite{Beltukov 2013}.

Using Eq.~(\ref{eq:cor}), the correlation matrix can be written as
\begin{eqnarray}
    C_{ij} = \frac{N}{K}\sum_k \left\langle A_{ik}A_{jk} \right\rangle.   \label{eq:Corr}
\end{eqnarray}
We intentionally add the summation over bonds $k$, which effectively averages over all bond positions in the system.

For the square random matrix $\hat{A}$ defined by Eq.~(\ref{eq:A_num}), the average $\langle A_{ik}A_{jk}\rangle$ over different realizations of the random numbers $\xi_{ij}$ is nonzero only if $i=k$ or $j=k$. Non-diagonal elements are $C_{ij}=-\Omega^2$ if $i$ and $j$ are neighbors and $C_{ij}=0$ otherwise. Diagonal elements are $C_{ii}=6\Omega^2$. Therefore, the structure of the matrix $\hat{A}$ defined by Eq.~(\ref{eq:A_num}) gives the same correlation matrix $\hat{C}$ as was considered in Section~\ref{sec:Wishart}.

For $\varkappa>0$ we can use two realizations of square random matrices defined by Eq.~(\ref{eq:A_num}): $\hat{A}^{(0)}$ and $\hat{A}^{(1)}$. The resulting rectangular matrix $\hat{A}$ can be obtained by inserting $\varkappa N$ columns of the matrix $\hat{A}^{(1)}$ into the matrix $\hat{A}^{(0)}$. This random insertion of the new columns corresponds to a random addition of new bonds to the vibrational system. This procedure results in the same correlation matrix $\hat{C}$.

In this section we have considered the nearest neighbor case with the unit radius of bonds $R=1$. The generalization of this model for arbitrary $R$ is given in Appendix~\ref{append:RMT}. We will show that increasing the radius $R$ leads to a rapid convergence to the vibrational properties obtained by the random matrix theory using the relation (\ref{eq:GG}). For $R=1$ there are only $n_{\rm nz}=7$ nonzero elements in each column of the matrix $\hat{A}$. For $R=2$ and $R=3$ this number is $n_{\rm nz}=33$ and $n_{\rm nz}=123$ respectively. The influence of the sparsity in the case of uncorrelated Wishart ensemble was discussed in~\cite{Beltukov 2011}. 

\section{Vibrational density of states}\label{sec:VDOS}

\begin{figure}[t]
    \centering
    \includegraphics[scale=0.65]{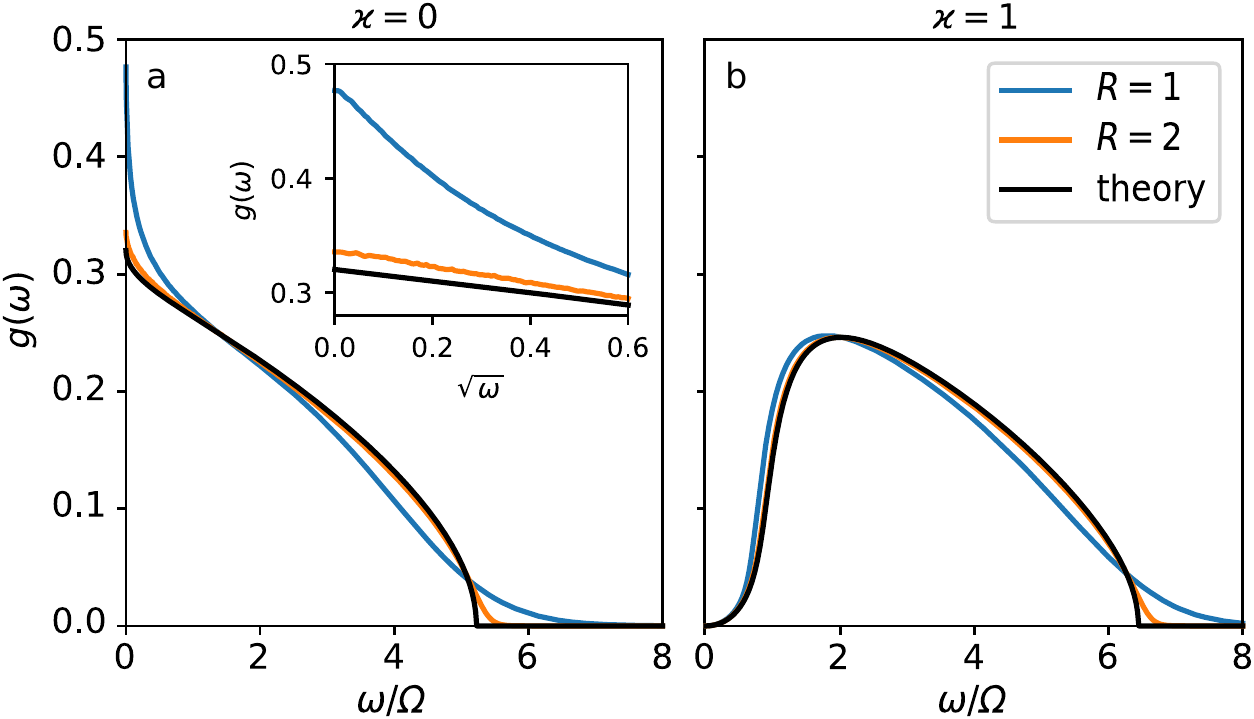}
    \caption{(Color online) The VDOS of the ensemble $\hat{M} = \hat{A}\hat{A}^T$ for $\varkappa=0$ and $\varkappa=1$. Black lines show the theoretical result obtained by Eqs.~(\ref{eq:watson})--(\ref{eq:CH}). Color lines show the numerical analysis of the system with $N=400^3$ atoms using the Kernel Polynomial Method for different radii of bonds: $R=1$ and $R=2$.}
    \label{fig:VDOS-KPM}
\end{figure}

To analyze the VDOS $g(\omega)$, we consider the normalized trace of $\hat{G}(z)$, which is the Stieltjes transform of $g(\omega)$:
\begin{gather}
     F(z) = \frac{1}{N}\Tr \hat{G}(z) = \int\frac{g(\omega)}{z-\omega^2}d\omega.   \label{eq:F(Z)}
\end{gather}
For regular correlation matrix $\hat{C}$, we can calculate a similar quantity $F_0(Z)=\frac{1}{N}\Tr \hat{G}_0(Z)$. Using the dispersion law for the cubic lattice (\ref{eq:disp}), we find
\begin{equation}
    F_0(Z) = \frac{1}{2\Omega^2}W_s\left(\frac{Z}{2\Omega^2}-3\right),  \label{eq:watson}
\end{equation}
where $W_s$ is the third Watson integral~\cite{Zucker 2011}. 
On the one hand, from Eq.~(\ref{eq:GG}) we know the relation $ZF_0(Z) = z F(z)$. On the other hand, we can express the VDOS as $g(\omega) = \frac{2\omega}{\pi}\Im F(\omega^2 - i0)$. As a result, we find
\begin{equation}
    g(\omega) = \frac{2\omega}{\pi}\Im \frac{1}{Z(\omega^2)},  \label{eq:g-closed}
\end{equation}
where the complex parameter $Z$ depends on the real parameter $\omega^2$ through the following complex equation
\begin{equation}
    \varkappa Z + Z^2F_0(Z) = \omega^2.   \label{eq:CH}
\end{equation}
This equation defines a contour on a complex plane, which is known as a critical horizon \cite{Burda 2006}. For a given parameter $\omega$, Eq.~(\ref{eq:CH}) has multiple solutions. We choose a physical one with $\Im Z(\omega^2) < 0$ which corresponds to $g(\omega)>0$.

\begin{figure}[t]
    \centering
    \includegraphics[scale=0.65]{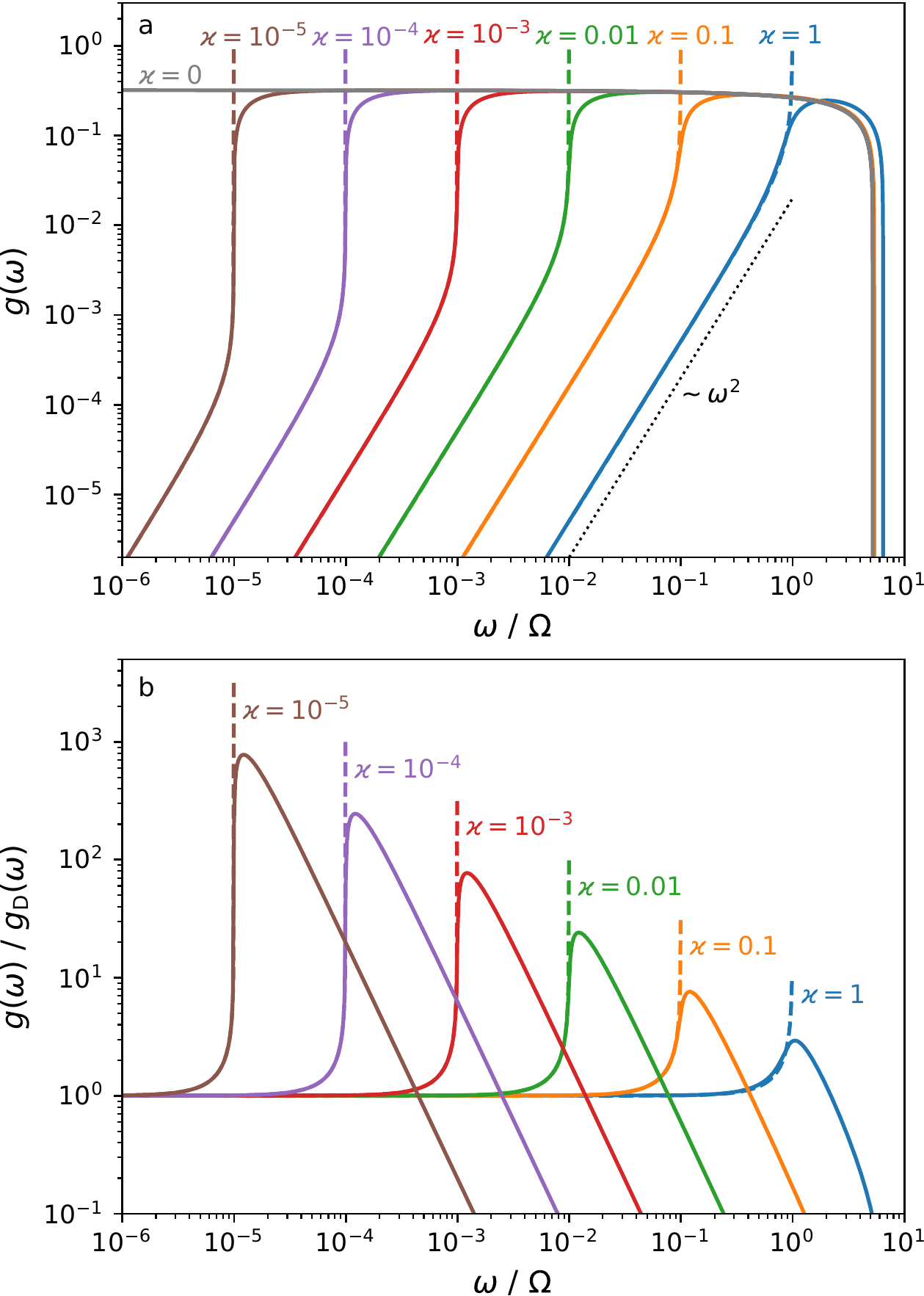}
    \caption{(Color online) a) The VDOS $g(\omega)$ for different values of the parameter $\varkappa$. Solid lines show the theoretical result obtained by Eqs.~(\ref{eq:watson})--(\ref{eq:CH}). For each value of $\varkappa$, the phononic VDOS $g_{\rm ph}(\omega)$ defined by Eq.~(\ref{eq:g_ph}) is shown by a dashed line. b) The same for the reduced VDOS $g(\omega)/g_\textsc{d}(\omega)$.}
    \label{fig:VDOS-BP}
\end{figure}

Equations (\ref{eq:watson})--(\ref{eq:CH}) defines the VDOS $g(\omega)$ in an implicit form, which can be solved numerically. The result is presented in Figs.~\ref{fig:VDOS-KPM} and~\ref{fig:VDOS-BP}. For $\varkappa > 0$, one can see a low-frequency range with the Debye law $g(\omega)\sim\omega^2$. However, for $\varkappa=0$, the VDOS has a constant low-frequency limit. Such behavior of the VDOS was observed in the random matrix model and the jamming transition~\cite{Beltukov 2013, OHern 2003}. The animated plot of the transition between a crystalline VDOS ($\varkappa = \infty$) and a soft amorphous
one ($\varkappa = 0$) is presented in Supplemental Materials~\cite{SM}.

Figure \ref{fig:VDOS-KPM} demonstrates a good agreement between the theory and the numerical VDOS calculated for a finite interaction radius $R$ for a system with $400^3$ atoms using the Kernel Polynomial Method~\cite{Beltukov 2018, Beltukov 2016 PRE, KPM 2006}. The results for $R=3$ is not shown in Fig.~\ref{fig:VDOS-KPM} because the difference with the theory is much smaller than the line thickness. Therefore, the theory is applicable for a finite interaction radius, which is important to describe amorphous solids. 

%It assumes the long-range interaction between atoms. However, the numerical calculations shows that sparse random matrices $\hat{A}$ with the same pair correlation matrix $\hat{C}$ have the same vibrational properties while the number of nonzero elements in each column of the matrix $\hat{A}$ is much bigger than one (see Section~\ref{sec:num}). For $R=2$, there are ... nonzero elements in each column of the matrix $A$ and the VDOS is close to the theoretical prediction (Fig.~\ref{fig:VDOS-KPM}). For $R=1$, there are only 7 such nonzero elements. In this case, the VDOS is slightly different from the theoretical prediction, which mostly concerns the high-frequency tail.

\section{Dynamical structure factor}\label{sec:DSF}

\begin{figure*}[t]
    \centering
    \includegraphics[scale=0.65]{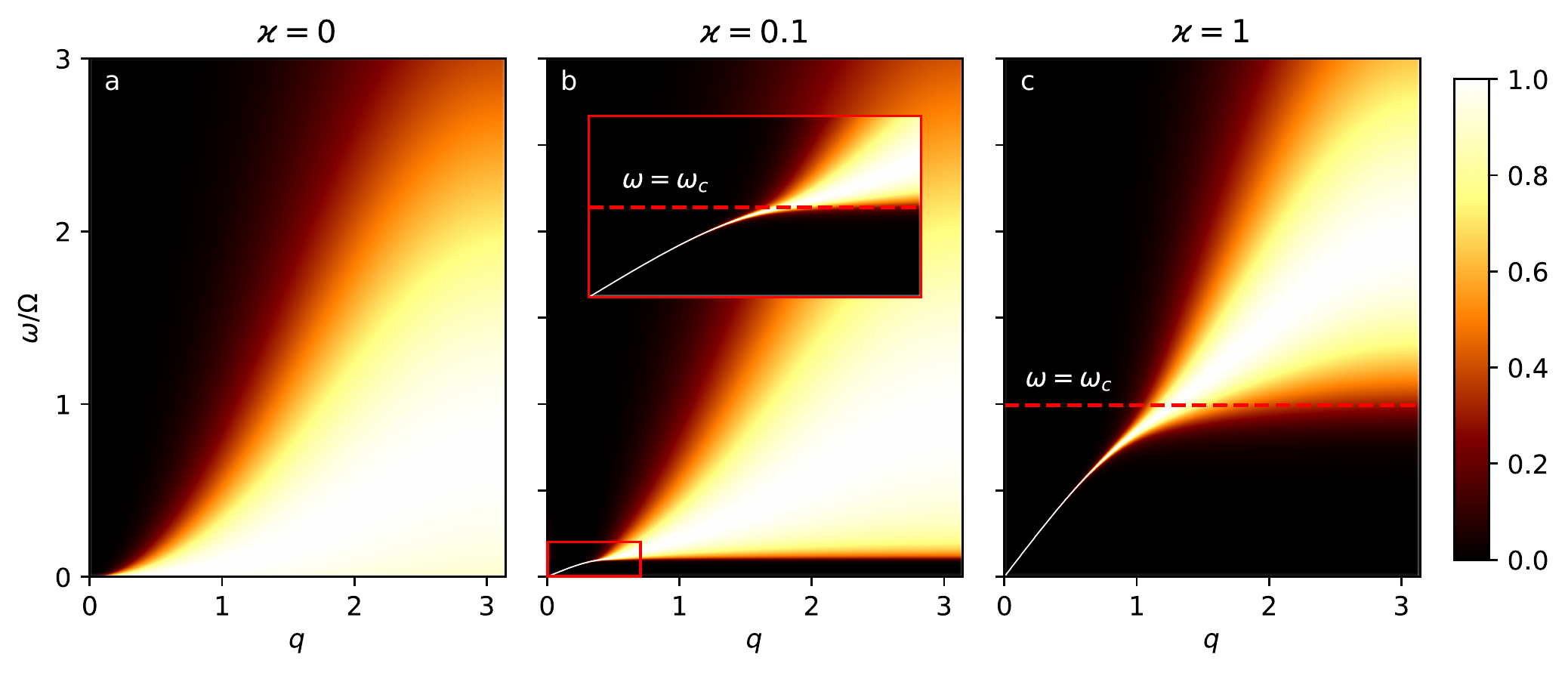}
    \caption{(Color online) The dynamical structure factor for $\varkappa=0$, $\varkappa=0.1$, and $\varkappa=1$. Color represents the normalized Fourier transform of eigenmodes ${\cal F}({\bf q}, \omega)/\max_\omega {\cal F}({\bf q}, \omega)$ for ${\bf q}$ along the direction [100] in the reciprocal space of the lattice. Horizontal dashed lines indicate the Ioffe-Regel frequency. Inset in panel (b) shows the magnification of the Ioffe-Regel crossover for $\varkappa=0.1$.}
    \label{fig:SF}
\end{figure*}

To analyze the spatial structure of the vibration modes, we calculate the DSF, which specifies the relation between the frequency $\omega$ and the wavevector ${\bf q}$~\cite{Tanaka 2008}. In the scalar model under consideration, the DFS has the form $S({\bf q}, \omega) = (k_BTq^2/m\omega^2) {\cal F}({\bf q}, \omega)$ with the Fourier transform of eigenvectors defined as 
\begin{equation}
    {\cal F}({\bf q}, \omega) = \sum_n \big| \langle n|{\bf q}\rangle\big|^2 \delta(\omega - \omega_n),
\end{equation}
where $\langle n|{\bf q}\rangle$ is a projection of $n$-th eigenmode to the plane wave with the wavevector ${\bf q}$. The DSF can be calculated from the resolvent:
\begin{equation}
    S({\bf q}, \omega) = \frac{2k_BTq^2}{\pi m \omega}\Im\langle{\bf
    q}|\hat{G}(\omega^2)|{\bf q}\rangle.
\end{equation}
Using the duality relation~(\ref{eq:GG}) and the dispersion law $\langle{\bf
    q}|\hat{G}_0(Z)|{\bf q}\rangle=1/(Z-\omega_0^2({\bf q}))$, the resulting dynamical structure factor can be presented in the form of the damped harmonic oscillator (DHO):
\begin{equation}
    S({\bf q}, \omega) =
    \frac{k_B T}{\pi m}\frac{2q^2\Gamma({\bf q}, \omega)}{(\omega^2-q^2E({\bf q}, \omega))^2 + \omega^2\Gamma^2({\bf q}, \omega)},   \label{eq:DHO}
\end{equation}
where the Young modulus is
\begin{gather}
    E({\bf q},\omega) = \frac{\omega_0^2({\bf q})}{q^2}\Re\frac{\omega^2}{Z(\omega^2)},   \label{eq:Young}
\end{gather}
and the damping is
\begin{equation}
    \Gamma({\bf q}, \omega) = \omega_0^2({\bf q}) \Im\frac{\omega}{Z(\omega^2)} = \frac{\pi}{2}\omega_0^2({\bf q}) g(\omega).   \label{eq:Gamma}
\end{equation}
Figure~\ref{fig:SF} shows the normalized Fourier transform of eigenvectors for different $\varkappa$.  This figure represents the shape of the dynamic structure factor and makes it possible to qualitatively determine the relation
between the frequency $\omega$ and the wave vector ${\bf q}$.  For $\varkappa=0$, there is no exact relation between the frequency $\omega$ and the wavevector ${\bf q}$. Such a broad DSF was attributed to \emph{diffusons} \cite{Allen 1999, Beltukov 2013}. For $\varkappa=1$, in the low-frequency range, there is a linear dispersion $\omega \sim {\bf q}$ with a small broadening due to a small scattering of plane waves. Such low-frequency vibrations are propagating \emph{phonons}. However, in the dominant frequency range, there is a broad behavior of the DSF. Thus, for nonzero $\varkappa$, there is a crossover between phonons and diffusons which is known as the {\em Ioffe-Regel crossover}. In this paper we do not consider the Anderson localization which affects only a small part of high-frequency vibrations~\cite{Allen 1999, Beltukov 2013}.

\section{Ioffe-Regel crossover}\label{sec:IRC}
To analyze the Ioffe-Regel crossover, we consider the low-frequency range $\omega \ll \Omega$. In this case we can use a small-argument expansion of $F_0(Z)$. For \emph{any} three-dimensional system with a linear dispersion $\omega_0({\bf q}) = \Omega q$ for $q\to 0$, this expansion has a form
\begin{equation}\label{eq:asymp}
    F_0(Z) = - \frac{a^2}{\Omega^2} + \frac{\sqrt{-Z}}{4\pi\Omega^3} + O(Z)
\end{equation}
where the dimensionless constant $a$ is determined by the integral over the first Brillouin zone
\begin{equation}
    a^2 = \frac{\Omega^2}{V_\bz} \int_\bz \frac{d{\bf q}}{\omega^2_0({\bf q})},
\end{equation}
where $V_\bz$ is the volume of the first Brillouin zone.
For the cubic lattice under consideration $a = \sqrt{w_s/2}$, where $w_s = (\sqrt{3} - 1)/(96\pi^3)\Gamma^2(\frac{1}{24})\Gamma^2(\frac{11}{24})\approx0.505462$ is the third Watson constant~\cite{Zucker 2011}. Using Eq.~(\ref{eq:asymp}), the critical horizon can be found explicitly for $\omega \ll \Omega$ using an iterative solution of Eq.~(\ref{eq:CH}) (see Appendix~\ref{append:Z(w)}):
\begin{equation}
    \frac{1}{Z(\omega^2)} = \frac{\varkappa}{2\omega^2} + \frac{1}{\omega}\sqrt{f(\omega)+
    \frac{i\omega/4\pi\Omega^3}{\sqrt{\varkappa/2 + \omega\sqrt{f(\omega)}}}},   \label{eq:Z(w)}
\end{equation}
where 
\begin{equation}
    f(\omega) = \frac{\varkappa^2}{4\omega^2} - \frac{a^2}{\Omega^2}.   \label{eq:f}
\end{equation}
The sign of $f(\omega)$ significantly changes the behavior of $Z(\omega^2)$. The corresponding crossover frequency 
\begin{equation}
    \omega_c = \varkappa\frac{\Omega}{2a}
\end{equation}
separates the frequency domain into two frequency ranges. Using the result (\ref{eq:Z(w)}), we analyze vibrational properties in both frequency ranges separately.

\subsection{Phonon frequency range}\label{sec:ph}

In this section we will show that the frequency range $\omega<\omega_c$ is occupied by phonons with well-defined dispersion relation $\omega(q)$. Using Eq.~(\ref{eq:Z(w)}) for $\omega<\omega_c - \delta$ with $\delta\ll\omega_c$, we obtain the VDOS
\begin{equation}
    g_{\rm ph}(\omega) = \frac{\omega}{4\pi^2(a\Omega)^{3/2}}\sqrt{\frac{\omega_c -\sqrt{\omega_c^2-\omega^2}}{\omega_c^2-\omega^2}}, \quad \omega < \omega_c.   \label{eq:g_ph}
\end{equation}
In the low-frequency range $\omega \ll \omega_c$, the obtained VDOS has the Debye behavior
\begin{equation}\label{eq:Debay}
    g_\textsc{d}(\omega) = \frac{\omega^2}{2\pi^2 v_0^3}
\end{equation}
with low-frequency phonon velocity $v_0=\Omega \sqrt{\varkappa}$, which corresponds to a static Young modulus $E_0 = \Omega^2\varkappa$. For $\varkappa = 0$ the Young modulus becomes zero, which means a soft system without propagation of phonons. Figure~\ref{fig:VDOS-BP}(b) demonstrates the boson peak in the reduced VDOS $g_{\rm ph}(\omega)/g_\textsc{d}(\omega)$ for different values of the parameter $\varkappa$. The height of the boson peak is proportional to $\varkappa^{-1/2}$, which diverges for $\varkappa\to 0$.

In this frequency range, the Young modulus $E({\bf q}, \omega)$ depends on the frequency only
\begin{equation}
    E_{\rm ph}(\omega) = \frac{\Omega^2\varkappa}{2}\left(1 + \sqrt{1-\frac{\omega^2}{\omega^2_c}}\right).   \label{eq:E(w)}
\end{equation}
At zero frequency we obtain the static Young modulus $E_{\rm ph}(0) = E_0 = \Omega^2\varkappa$. Using Eq.~(\ref{eq:E(w)}) and the relation $\omega^2/q^2 = E_{\rm ph}(\omega)$, we obtain the dispersion of phonons
\begin{equation}
    \omega(q) = \Omega aq\sqrt{2q_c^2-q^2},
\end{equation}
where the crossover wavenumber $q_c = \sqrt{\varkappa/2a^2}$ corresponds to the crossover frequency $\omega_c$. For low-frequency modes with $q\ll q_c$, the dispersion is linear: $\omega(q) = v_0q$. The group velocity has a form
\begin{equation}
    v_g(q) = 2a\Omega\frac{q_c^2 - q^2}{\sqrt{2q_c^2-q^2}}.
\end{equation}
For $q=q_c$ the group velocity of phonons becomes zero, which explained the divergence of the phononic VDOS for $\omega=\omega_c$.

\begin{figure}[t]
    \centering
    \includegraphics[scale=0.65]{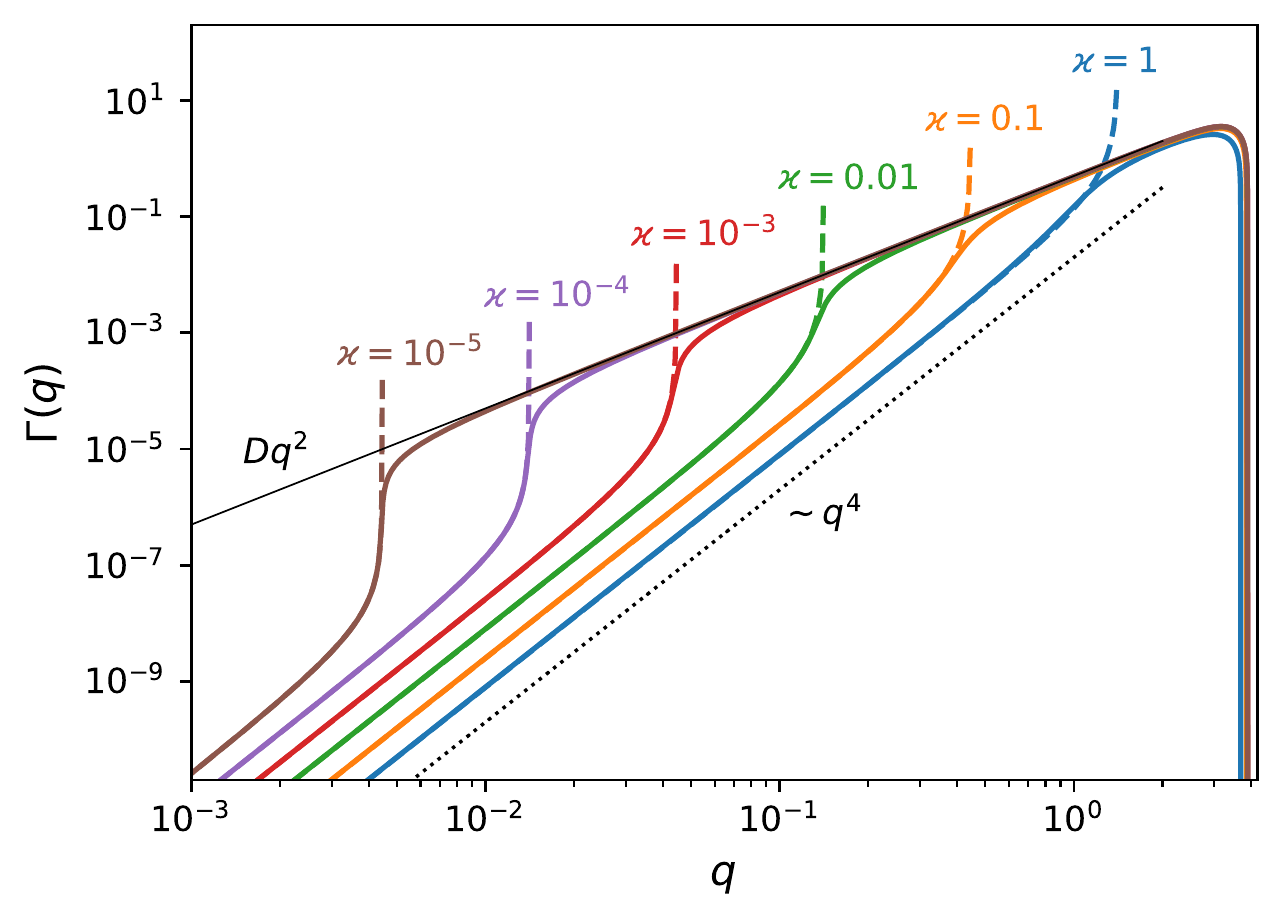}
    \caption{(Color online) The damping $\Gamma$ as a function of the wavevector ${\bf q}$ for different values of the parameter $\varkappa$. Solid lines show the theoretical result obtained by Eqs.~(\ref{eq:watson})--(\ref{eq:CH}), (\ref{eq:Gamma}). For each value of $\varkappa$, the damping of phonons $\Gamma_{\rm ph}(q)$ defined by Eq.~(\ref{eq:Gamma_ph}) is shown by a dashed line. Thin solid black line shows the diffusion law $\Gamma = D q^2$ with $D=a\Omega$. Dotted line shows the Rayleigh scattering $\gamma\sim q^4$. }
    \label{fig:Gamma}
\end{figure}

According to Eq.~(\ref{eq:Gamma}), the damping $\Gamma({\bf q},\omega)$ follows the vibrational density of states $g(\omega)$, which can be written using the dispersion relation:
\begin{equation}
    \Gamma_{\rm ph}(q) = \frac{\Omega q^4}{8\pi a}\frac{\sqrt{2q_c^2-q^2}}{q_c^2-q^2}.  \label{eq:Gamma_ph}
\end{equation}
For low-frequency modes with $q\ll q_c$, the damping $\Gamma_{\rm ph}({\bf q}) \sim q^4$, which corresponds to the Rayleigh scattering from disorder (Fig.~\ref{fig:Gamma}). In amorphous bodies, additional resonant scattering of phonons by quasilocal vibrations can occur~\cite{Buchenau 1992}. The number of quasilocal vibrations decreases with increasing relaxation time~\cite{Rainone 2020}, and this phenomenon goes beyond the general assumptions (i) and (ii) given in the introduction. However, the quasilocal vibrations can be studied in the framework of the random matrix model, which will be discussed in Section~\ref{sec:QLV}.

\begin{figure}[t]
    \centering
    \includegraphics[scale=0.65]{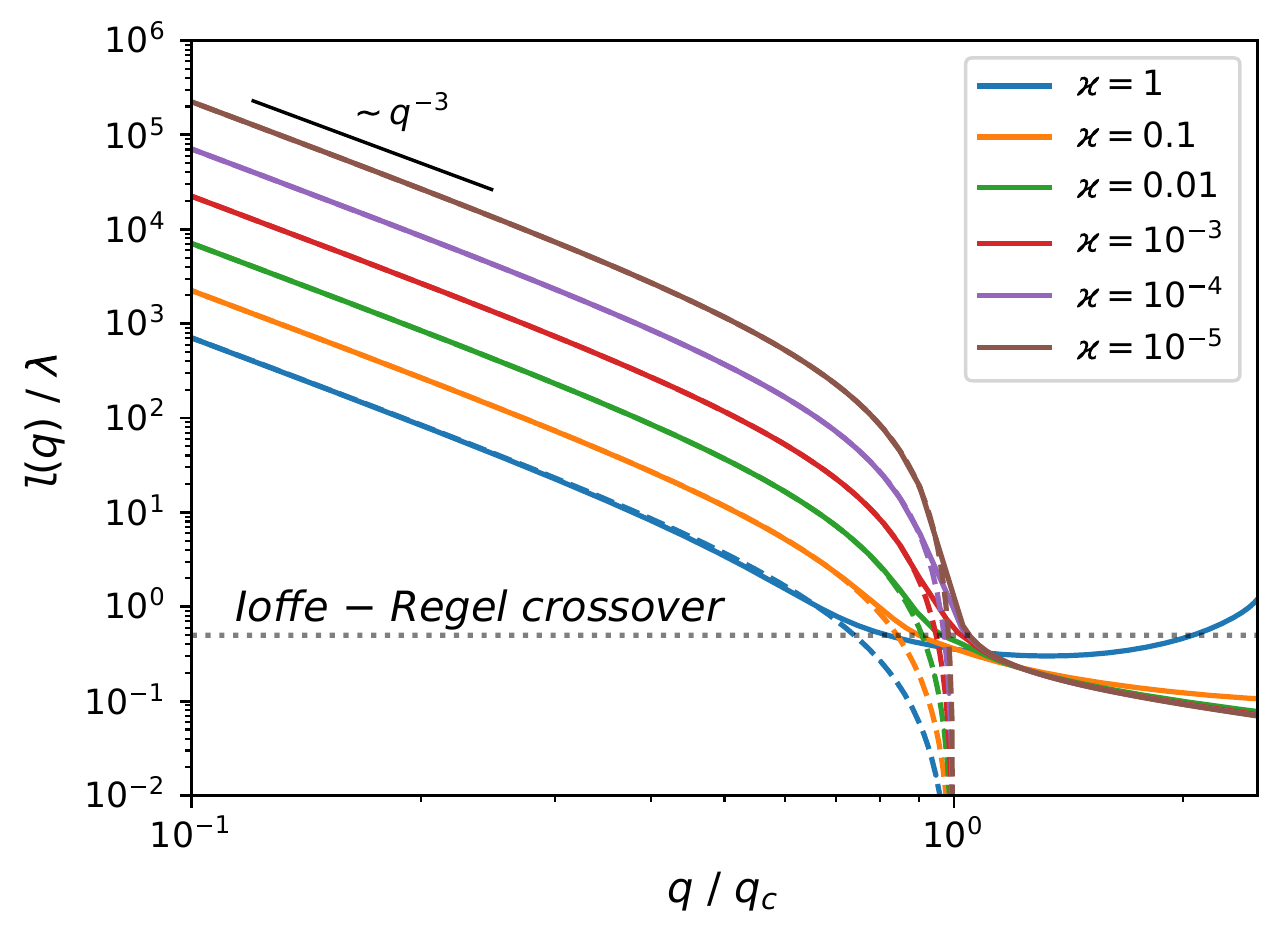}
    \caption{(Color online) The ratio of the mean free path $l$ to the wavelength $\lambda$ as a function of the reduced wave vector $q/q_c$ for different values of the parameter $\varkappa$. Solid lines show the theoretical result obtained by Eqs.~(\ref{eq:watson})--(\ref{eq:CH}), (\ref{eq:Young}), (\ref{eq:Gamma}). Dashed lines show the result for phonons using $l_{\rm ph}(q)$ defined by Eq.~(\ref{eq:l_ph}). Horizontal dotted line marks the Ioffe-Regel criterion $l\approx \lambda/2$.} 
    \label{fig:IR}
\end{figure}

The mean free path of phonons is defined by the ratio of the group velocity $v_g(q) = d \omega(q)/d q$ and the damping $\Gamma_{\rm ph}(q)$:
\begin{equation}
    l_{\rm ph}(q) = \frac{v_g(q)}{\Gamma_{\rm ph}(q)} = \frac{16\pi a^2}{ q^4}\frac{(q_c^2-q^2)^2}{2q_c^2-q^2}.   \label{eq:l_ph}
\end{equation}
The mean free path $l_{\rm ph}(q)$ becomes of the order of the wavelength $\lambda=2\pi/q$ near the crossover frequency $\omega_c$ (Fig.~\ref{fig:IR}).  It means, that the frequency $\omega_c$ defines the Ioffe-Regel crossover, which can be written as $l_{\rm ph}(q)/\lambda\approx1/2$~\cite{Beltukov 2013}.

For an arbitrary relation between $\omega$ and $\omega_c$, we can also find the dispersion law $\omega(q)$ using the definition $\omega^2/q^2 = E(\omega, {\bf q})$ with $E(\omega, {\bf q})$ defined by Eq.~(\ref{eq:Young}). The resulting ratio $l(q)/\lambda$ is shown in Fig.~\ref{fig:IR} by solid lines. However, the notion of dispersion law $\omega(q)$ for $\omega>\omega_c$ is not meaningful because the damping greatly exceeds the frequency in this case. Indeed, the DSF demonstrates a broad shape for $\omega>\omega_c$ (Fig.~\ref{fig:SF}).

\subsection{Diffuson frequency range}\label{sec:diff}

In this section we study vibrational properties in the frequency range $\omega > \omega_c$ in more detail. From Eq.~(\ref{eq:Z(w)}) for $\omega > \omega_c + \delta$ with $\delta\ll\omega_c$, we obtain the VDOS in the form
\begin{equation}
    g_{\rm diff}(\omega) = \frac{2a}{\pi\omega\Omega}\sqrt{\omega^2-\omega_c^2}, \quad \omega > \omega_c.  \label{eq:g_diff}
\end{equation}
This form is similar to the form of the Marchenko-Pastur law obtained for the uncorrelated Wishart ensemble~\cite{Beltukov 2015}.

Figure~\ref{fig:VDOS-BP}(b) shows that the frequency of the boson peak $\omega_b$ is slightly bigger than the crossover frequency $\omega_c$. For $\varkappa\ll 1$ we obtain the relation $\omega_b = \sqrt{3/2}\omega_c$ from Eq.~(\ref{eq:g_diff}). As a result, the Young modulus $E_0$ is proportional to the boson peak frequency $\omega_b$. This relation was observed by other experimental and theoretical groups~\cite{Kojima 2011, Vitelli-2010}.

If $\varkappa=0$, the number of degrees of freedom $N$ is equal to the number of bonds $K$. In the jamming transition, this state is known as the isostatic state. In this case, the macroscopic rigidity becomes zero and the low-frequency VDOS does not follow the Debye law because $\omega_c=0$. From Eq.~(\ref{eq:g_diff}) for $\omega_c=0$, we obtain $g_{\rm is}(0) = 2a/\pi\Omega$.
Inset in Fig.~\ref{fig:VDOS-KPM}(a) shows that isostatic VDOS in the numerical random matrix model also has a finite zero-frequency limit. It is worth to note that the isostatic VDOS in the inset has a linear form as a function of $\sqrt{\omega}$, which means the low-frequency cusp-like singularity in the linear scale. Such a form of the isostatic VDOS was observed numerically in the random matrix model~\cite{Beltukov 2015} and the jamming transition~\cite{OHern 2003, Wyart 2005, Narayan 2020}. We will show that this behavior is related to the diffusive nature of low-frequency vibrations in the isostatic case. To demonstrate this, we consider the isostatic DSF $S_{\rm is}({\bf q}, \omega) = (k_BTq^2/m\omega^2) {\cal F}_{\rm is}({\bf q}, \omega)$, which is determined by the Fourier transform of the eigenmodes in the form (see Eq.~(\ref{eq:DHO}) for $\varkappa=0$)
\begin{equation}
    \mathcal{F}_{\rm is}({\bf q}, \omega) = \frac{1}{\pi}\frac{2\Gamma({\bf q}, \omega)}{\omega^2 + \Gamma^2({\bf q}, \omega)}.   \label{eq:F_is}
\end{equation}
From Eq.~(\ref{eq:Gamma}) for the isostatic case, we obtain $\Gamma({\bf q},\omega) = \frac{a}{\Omega}\omega_0^2({\bf q})$.
The obtained $\mathcal{F}_{\rm is}({\bf q}, \omega)$ coincide with the Fourier transform of the random walk on a lattice~\cite{Beltukov 2013}. Thus, we analytically confirm the idea that diffusons in the isostatic case can be considered as random walks of atomic displacements~\cite{Beltukov 2013}. In the continuous limit $q\ll1$, $\Gamma({\bf q}, \omega) = Dq^2$ with diffusivity $D=a\Omega$. 

The VDOS is related to the Fourier transform of the eigenmodes as
\begin{equation}
    g_{\rm is}(\omega) = \frac{1}{V_\bz}\int_\bz \mathcal{F}_{\rm is}(\mathbf{q}, \omega)d\mathbf{q}.
\end{equation}
Since $\mathcal{F}_{\rm is}(\mathbf{q}, \omega)$ depends on $\Gamma({\bf q}, \omega)$, which depends on $g_{\rm is}(\omega)$, we obtain the observed singular behavior of the isostatic VDOS:
\begin{equation}
    g_{\rm is}(\omega) \simeq \frac{2a}{\pi\Omega} - \frac{1}{4\pi^2}\sqrt{\frac{\omega}{2a^3\Omega^3}}.   \label{eq:g_is}
\end{equation}

For nonzero $\varkappa$, there is a nonzero frequency-dependent Young modulus 
\begin{equation}
    E_{\rm diff}(\omega) = \frac{\Omega^2\varkappa}{2} +
    \frac{1}{8\pi}\sqrt{\frac{\omega^3 \Omega}{2a^3}}.
\end{equation}
However, in the dominant part of the diffuson frequency range $\omega_c \ll \omega \ll \Omega$, the existence of nonzero Young modulus is negligible and the DSF is described by the isostatic one: 
\begin{equation}
    S_{\rm diff}({\bf q}, \omega) = \frac{k_B T q^2}{\pi m \omega^2}\frac{2\Gamma({\bf q}, \omega)}{\omega^2 + \Gamma^2({\bf q}, \omega)},    \label{eq:diffusons}
\end{equation}
which verifies the notion of diffusons introduced in~\cite{Allen 1993, Allen 1999}. Figure~\ref{fig:Gamma} shows a crossover between the low-frequency Rayleigh scattering $\Gamma\propto q^4$ and the diffusion damping $\Gamma\propto q^2$. Such a quadratic dependence above the Ioffe-Regel crossover was observed experimentally~\cite{Sette-1998, Ruocco-2001, Christie-2007}. The crossover between $\Gamma\propto q^4$ and $\Gamma\propto q^2$ was observed experimentally~\cite{Ruffle 2006, Monaco 2009a} and in the molecular dynamics simulations~\cite{Monaco 2009}.

\section{Quasilocalized vibrations}\label{sec:QLV}

The sparsity of the matrix $\hat{A}$ results in some deviation of the vibrational properties due to the non-Gaussian statistical properties of the random matrix ensemble. Therefore, we will study additional vibrational properties of the numerical random matrix model for a finite radius of bonds $R$.

%One can assume that in real amorphous bodies there are soft-local sites which lead to additional resonant scattering of phonons by quasilocal vibrations. Statistics of these soft-local areas are beyond the scope of consideration approximation of pair correlations. Wherein existence of such areas lead to the appearance of additional low-frequency (soft) modes in the region $\omega < \omega_c$ and blurs of the vibrational density states step in the crossover region.

\begin{figure}[t]
    \includegraphics[scale=0.65]{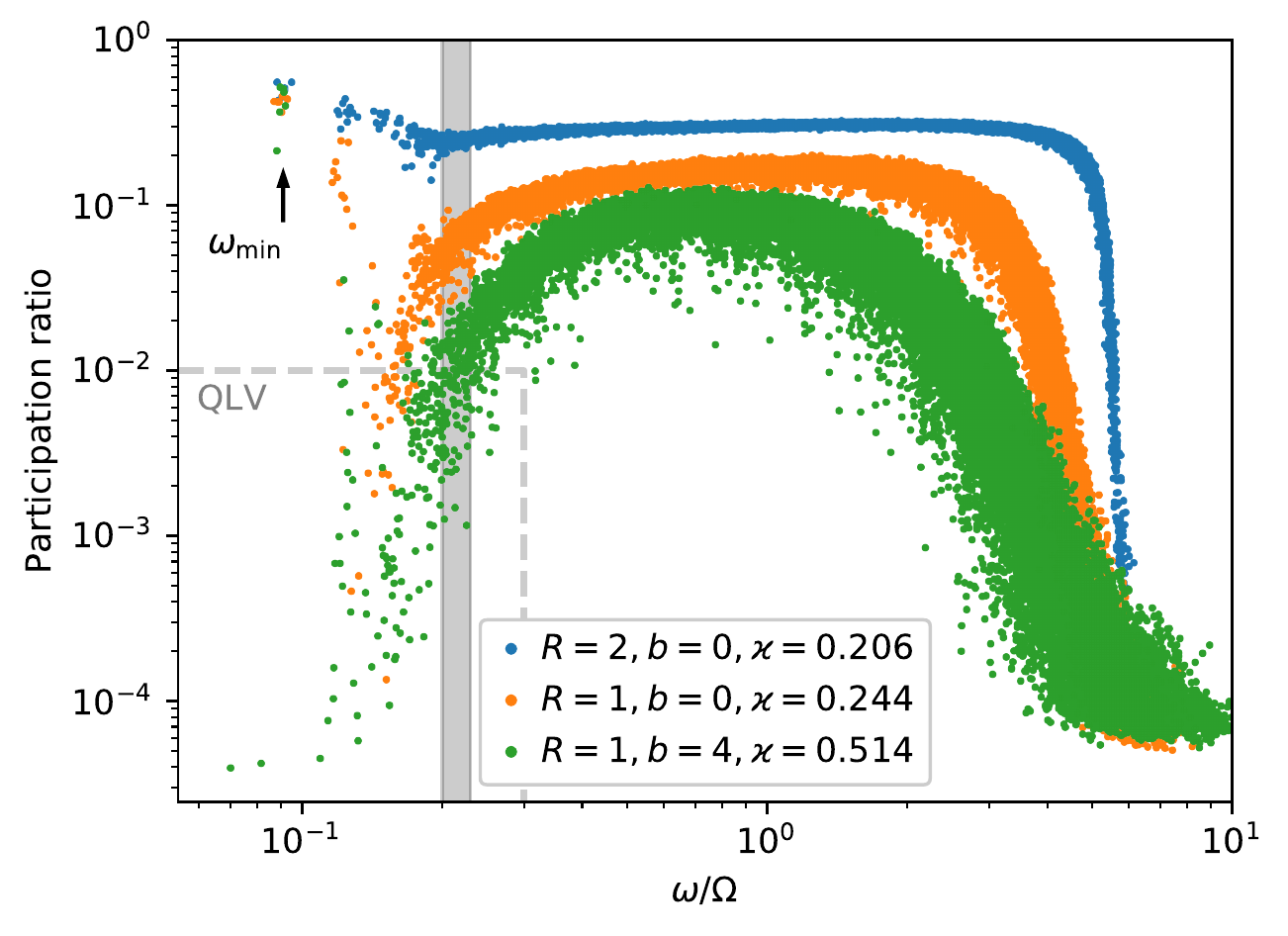}
    \caption{(Color online) Participation ratio $P$ for different values of the bond's radius $R$ and the non-Gaussian parameter $b$. The value of the parameter $\varkappa$ is chosen to hold the static Young modulus $E_0=0.2\Omega^2$. In all cases the system size is $N=30^3$. The gray vertical strip indicates the position of the boson peaks. The arrow shows the minimal phonon frequency $\omega_{\rm min}$. Gray dashed line roughly outlines the region of quasilocalized vibrations.}
    \label{fig:PR}
\end{figure}

The presence of the quasilocalized modes can be studied using the participation ratio
\begin{equation}
    P(\omega_n) = \frac{1}{N\sum_i \langle n|{\bf r}_i\rangle^4},
\end{equation}
where $\langle n|{\bf r}_i\rangle$ is the projection of $n$-th eigenmode to $i$-th site. For delocalized modes (phonons and diffusons) $P\sim 1$, while for localized ones $P\sim 1/N$. For a small system, the quasilocalized modes are weakly entangled with phonons due to a small number of phonons in such a case. The participation ratio clearly shows the presence of quasilocalized modes for $R=1$: there are vibrations with $P\ll 1$ between phonons, which have $P\sim 1$ (Fig.~\ref{fig:PR}).
    
For bigger values of $R$, we have shown that the VDOS becomes close to the prediction of the random matrix theory (Fig.~\ref{fig:VDOS-KPM}). The participation ratio for $R=2$ did not show any vibrational mode, which can be clearly identified as quasilocal vibration (Fig.~\ref{fig:PR}). Thus, the increasing of the interaction radius reduces the sparsity of the matrices $\hat{A}$ and $\hat{M}$, which decreases the number of quasilocalized modes. 

For a fixed value of the parameter $\varkappa$, the values of the static Young modulus $E_0$, the boson peak frequency $\omega_b$, and Ioffe-Regel frequency $\omega_\ir$ slightly depend on the radius of bonds $R$. For better visual performance, the parameter $\varkappa$ is chosen to hold the same static Young modulus $E_0=0.2\Omega^2$ in Fig.~\ref{fig:PR}. In this case the boson peak frequency $\omega_b$ stays within a narrow frequency range shown by the vertical strip in Fig.~\ref{fig:PR}. The Ioffe-Regel frequency is also close to this frequency range.

There are different possibilities to increase the sparsity of the interaction network more than it is possible for the smallest bond radius $R=1$. The first one is to consider the lattice with a smaller number of neighbors (e.g. diamond lattice). However, it is a non-universal approach with a fixed sparsity of the matrices. The second one is to randomly put some random values to zero. However, it results in a loose network with a number of small independent clusters of atoms. The third one is to consider a non-Gaussian distribution of nonzero entries in the sparse random matrix $A$. Indeed, some of the random elements can be relatively small, which represents a weaker interaction between some degrees of freedom.

We settled on the third option and introduce the non-Gaussian parameter $b$ which increases the probability to obtain a small-magnitude interaction (see Appendix \ref{sec:exp}). The value $b=0$ corresponds to the Gaussian distribution, which was studied before. The participation ratio for $b=4$ is shown in Fig.~\ref{fig:PR}. The number of quasilocalized modes with $P\ll 1$ is significantly bigger in this case. There are some quasilocalized modes, which frequency is smaller than the smallest phonon frequency $\omega_{\rm min}\approx 2\pi v_0/L$ in the periodic system of size $L=30$ (marked by the arrow in Fig.~\ref{fig:PR}).

\begin{figure}[t]
    \includegraphics[scale=0.65]{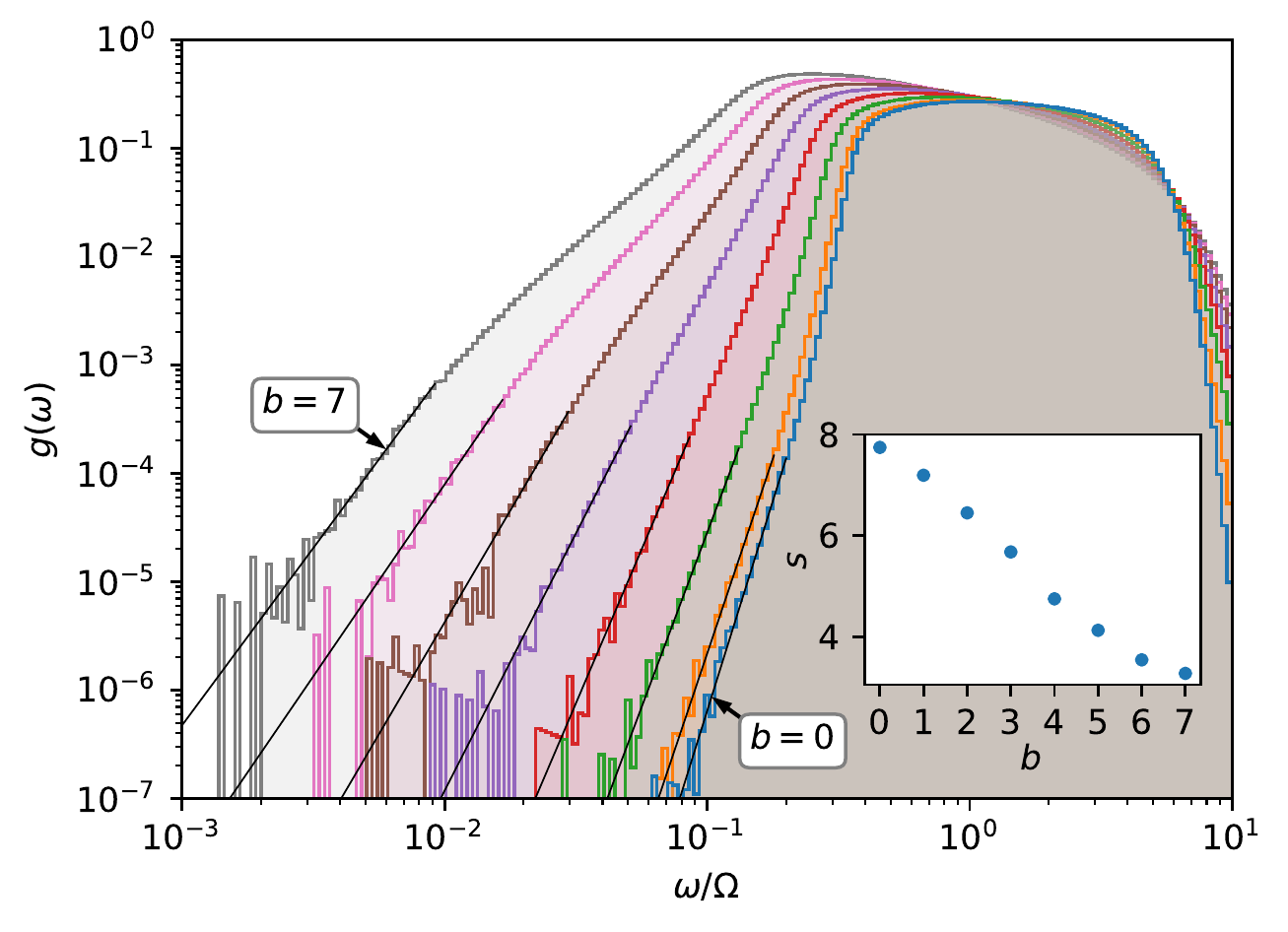}
    \caption{(Color online) The VDOS for a small system with $N=7^3$ atoms for $\varkappa=0.5$ and integer values of the non-Gaussian parameter $b$ from $b=0$ to $b=7$. The bond's radius is $R=1$. Each histogram was averaged over $6\cdot 10^6$ different realizations of the system. Thin lines show the fit to the power-law dependence $g_{\rm qlv}(\omega)\sim\omega^s$. Inset shows the dependence of the fitted value of $s$ on the non-Gaussian parameter $b$. }
    \label{fig:small}
\end{figure}

To study the distribution of the quasilocalized modes, we calculate the VDOS using the full diagonalization of small matrices and plot the histogram in Fig.~\ref{fig:small}. The smallest frequency of phonons $\omega_{\rm min}$ is limited by the system size. At the same time, there is no such constraint for quasilocalized modes~\cite{}. In this case the low-frequency range of the calculated VDOS represents the density of quasilocalized modes $g_{\rm qlv}(\omega)$. 

Figure~\ref{fig:small} shows the VDOS calculated for a small system with $N=7^3$ atoms for different values of the non-Gaussian parameter $b$. Increasing the value of $b$ leads to a substantial growth of the density of quasilocalized modes $g_{\rm qlv}$ (up to 6 orders of magnitude). In the low-frequency range, we observe the power-law dependence $g_{\rm qlv} \sim \omega^s$. The exponent $s$ was fitted using the maximum likelihood method assuming that the height of each bin in the histogram has the Poisson distribution. The increasing of the parameter $b$ decreases the power $s$ (see inset in Fig.~\ref{fig:small}). 

Figure~\ref{fig:small} was calculated using a system with $N=7^3$ atoms and $\varkappa=0.5$. The same power-law dependence with the same exponents $s$ was observed for $N=5^3$ and $N=10^3$. For a smaller value of the parameter $\varkappa = 0.2$, we observe similar values of the exponent $s$. In the case $R=2$, the non-Gaussian parameter $b$ also leads to the presence of the quasilocalized modes but for higher values of $b\sim 40$.

\section{Discussion}\label{sec:Disc}

We have shown that the random matrix approach can be applied numerically and analytically to study the general vibrational properties of amorphous solids like the boson peak and the Ioffe-Regel crossover. 

In this paper we obtain the vibrational properties in different analytical forms. The system of Eqs.~(\ref{eq:watson})--(\ref{eq:CH}) represents the precise form of the critical horizon $Z(\omega^2)$ for the simple cubic lattice with random bonds. Using Eqs.~(\ref{eq:g-closed}), (\ref{eq:DHO}), (\ref{eq:Young}), (\ref{eq:Gamma}), we find the vibrational properties without any approximation, which are shown by solid lines in Figs.~\ref{fig:VDOS-BP}--\ref{fig:IR}.

Then we apply the approximation $\omega\ll \Omega$ to find the exact analytical form of $Z(\omega^2)$ for any atomic ordering (Eq.~(\ref{eq:Z(w)})). We have checked that the result is almost indistinguishable from the exact solution of the simple cubic lattice for frequencies up to $\omega \approx 2\Omega$. 

Then we find vibrational properties in a simple analytical form for phonons  (Section~\ref{sec:ph}) and diffusons (Section~\ref{sec:diff}) separately. The obtained analytical results significantly expands the results of~\cite{Beltukov 2013}, which was obtained numerically. In the present paper we also introduce the notion of bonds as a many-body positive-definite interaction, which plays a crucial role in the theory.

The relation between the number of degrees of freedom in the system $N$ and the number of bonds $K$ plays a crucial role in many theoretical approaches like the effective medium theory (EMT)~\cite{Wyart 2005, Wyart 2010, DeGiuli 2014, DeGiuli 2015} and jammed solids~\cite{OHern 2003, Beltukov 2015}. The obtained scaling relations correspond to transverse vibrational properties of the jammed solids if we put $\varkappa\sim \Delta\phi^{1/2}$ and $\Omega\sim\Delta\phi^{(\alpha-2)/2}$~\cite{OHern 2003, Vitelli-2010}.

In terms of scaling relations, our results have a good agreement with the results of the EMT if we put $\varkappa\sim z-z_0$~\cite{Wyart 2010, DeGiuli 2014, DeGiuli 2015}. In particular, the relation $\omega_b \simeq \omega_\ir$ between boson peak frequency $\omega_b$ and Ioffe-Regel crossover frequency $\omega_\ir = \omega_c$ was also obtained in works~\cite{Wyart 2005, Wyart 2010} with the scaling-relations $\omega_b \sim \varkappa$ and $\omega_\ir \sim \varkappa$, which are also verified by numerical simulation~\cite{Mizuno 2017, Mizuno 2018}. The height of the boson peak is scaled as $g(\omega_b)/g_\textsc{d}(\omega_b) \propto \varkappa^{-1/2}$, which is also one of the predictions of the work~\cite{Wyart 2010}. 

However, there are original results obtained by the random matrix approach. As we mentioned above, we present the vibrational properties in different analytical forms, depending on whether we apply some approximations or not. We also explain the cusp-like singularity of the isostatic VDOS~\cite{Beltukov 2015, OHern 2003, Wyart 2005, Narayan 2020} using the notion of diffusons in this frequency range.

The random matrix approach can be also applied to analyze two-dimensional amorphous systems~\cite{Conyuh 2020}. It was shown that the height of the boson peak has logarithmic scaling with $g(\omega_b)/g_\textsc{d}(\omega_b) \propto {\rm log}(\varkappa^{-1})$ in this case. 

We also present a numerical model with finite interaction radius and the same pairwise correlation matrix $\hat{C}$. Starting from the radius of bonds $R=2$, there is almost perfect agreement between the theory and the numerical results. 

For small interaction radius $R=1$, there is some deviation of the vibrational properties due to non-Gaussian statistics of the matrix $\hat{A}$. We can enhance this deviation by introducing nonzero non-Gaussian parameter $b$. We observe the quasilocalized modes with the low-frequency VDOS $g_{\rm qlv}\sim\omega^s$. For $b=5$ and $R=1$ we obtain $s\approx 4$ (see inset in Fig.~\ref{fig:small}), which corresponds to the distribution $g_{\rm qlv}\sim\omega^4$, which was usually observed in amorphous solids~\cite{Lerner 2016, Mizuno 2017, Shimada 2018, Lerner 2017, Kapteijns 2018, Wang 2019}. Also, the power $s=3$ can be sometimes observed~\cite{Lerner 2017}. 

The non-Gaussian statistics can be considered as an additional inhomogeneity of amorphous system obtained with a finite relaxation time. We assume, that the equilibration of amorphous systems is a complicated process, which may lead to specific non-Gaussian properties of the near-equilibrium dynamical matrix. The present theory is essentially based on the stability properties of the dynamical matrix after the equilibration. At the same time, the existing theories investigate the stabilization and atomic rearrangement of unstable modes and the marginal stability of amorphous solids~\cite{Gurevich 2003, Mizuno 2017, Shimada 2020, Charbonneau 2016, Ji 2019}. In terms of the random matrix approach, the equilibration process leads to specific statistical properties of the matrix $\hat{A}$ which will be investigated in detail in future research.

Increasing the parameter $b$ increases the number of the quasilocalized modes, which results in the increasing of the VDOS $g(\omega)$ below the Ioffe-Regel crossover. Due to the resonance scattering of phonons on the quasilocalized vibrations, the damping $\Gamma(q)$ will be also increased below the Ioffe-Regel crossover. For a large number of the quasilocalized modes, the step-wise behavior of $g(\omega)$ and $\Gamma(q)$ near the Ioffe-Regel crossover will be smoothed out.

\section{Conclusion}\label{sec:Final}

We have shown that the random matrix theory can be applied numerically and analytically to study the general vibrational properties of amorphous solids like the boson peak and the Ioffe-Regel crossover.

To summarize, we present the correlated Wishart ensemble the random matrix approach, which takes into account only the most important correlations of random matrices, which ensure the mechanical stability (i) and the translation invariance (ii). In the framework of the random matrix approach, we find the vibrational density of states and the dynamical structure factor. We demonstrate the presence of the Ioffe-Regel crossover between low-frequency propagating phonons and diffusons at higher frequencies. The boson peak essentially appears near the Ioffe-Regel crossover. 

Using the numerical random matrix model, we verify the obtained results for sparse matrix $\hat{A}$ with $n_{\rm nz}\gg 1$ nonzero elements in each column. The presence of the non-Gaussian properties for small number $n_{\rm nz}$ leads to the presence of quasilocalized vibrations in the system. The introduced non-Gaussian parameter $b$ enhances the number of quasilocalized modes in the system. The quasilocalized vibrations make an additional contribution to the boson peak, which may be significant for large values of $b$. However, we demonstrate the presence of the boson peak and the Ioffe-Regel crossover without quasilocalized vibrations.

\section{Acknowledgments}
We wish to acknowledge D.\,A.~Parshin and V.\,I.~Kozub for valuable discussions. The authors thank the Council of the President of the Russian Federation for State Support of Young Scientists and Leading Scientific Schools (project no. MK-3052.2019.2) for the financial support.

\appendix

\section{The numerical model with an arbitrary radius of bonds}\label{append:RMT}

\begin{figure}
    \centering
    \includegraphics[scale=0.45]{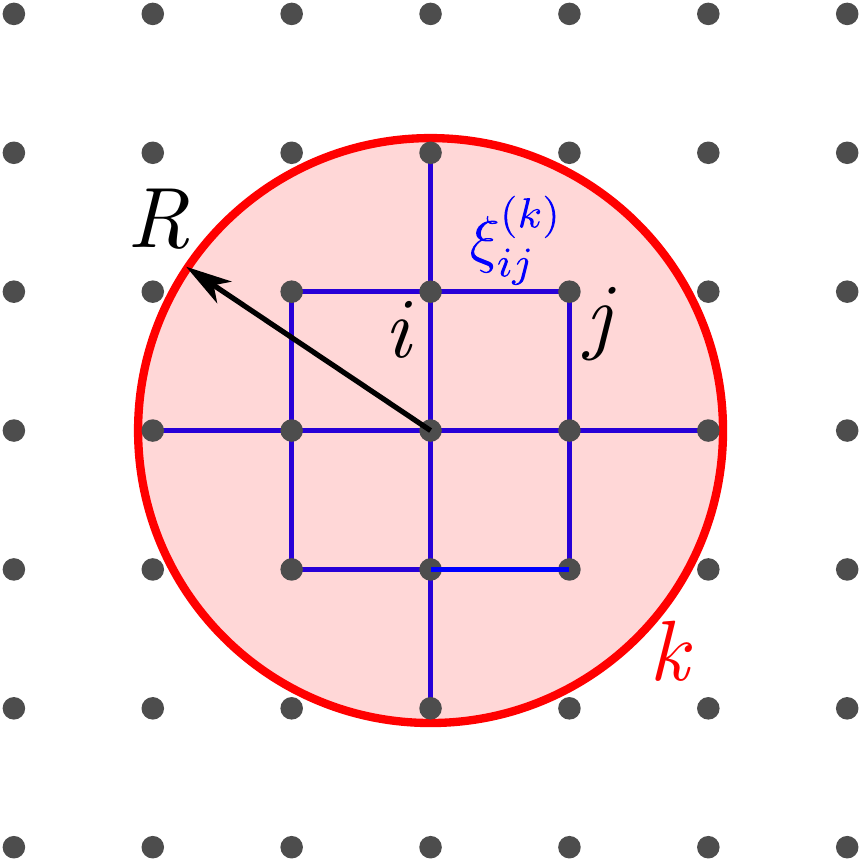}
    \caption{(Color online) A two-dimensional illustration of a bond $k$ with radius $R=2$. The nonzero elements $h_{ij}^{(k)}$ are indicated by straight lines between two atoms. Each nonzero $h_{ij}^{(k)}$ corresponds to the presence of a random interaction $\xi_{ij}^{(k)}$ between atoms $i$ and $k$ in the bond $k$.}
    \label{fig:bonds}
\end{figure}

To obtain the random matrix $\hat{A}$ for an arbitrary radius of bonds $R$ with a given correlation matrix $\hat{C}$, we use the following form of the matrix~$\hat{A}$:
\begin{equation}
    A_{ik} = \Omega\sqrt{\frac{d}{\cal N}}\sum_{j} h_{ij}^{(k)}\xi_{ij}^{(k)}.  \label{eq:A_R}
\end{equation}
Here $\xi_{ij}^{(k)}$ are Gaussian random numbers with unit variance and zero mean, which specifies the interaction between atoms $i$ and $j$ in the bond $k$. These random numbers are independent except the antisymmetry rule $\xi_{ij}^{(k)} = -\xi_{ji}^{(k)}$. The range of summation is determined by the mask $h_{ij}^{(k)}$, which is 1 if atoms $i$ and $j$ are neighbors with $r_{ik} \leq R$ and $r_{jk} \leq R$. Otherwise, $h_{ij}^{(k)}=0$. The prefactor in Eq.~(\ref{eq:A_R}) is determined by the number of dimensions $d=3$ and the mask size ${\cal N} = \sum_{ik} h_{ij}^{(k)}$, which is assumed to be the same for each bond $k$.

In the model under consideration, we have $K = {(1 + \varkappa)N}$ statistically equivalent bonds. We place $N$ of them uniformly in the system with $i$-th bond centered at $i$-th atom. Then we take $\varkappa N$ atoms randomly and place an additional set of $\varkappa N$ bonds around them. Using Eq.~(\ref{eq:Corr}), we obtain the correlation matrix
\begin{equation}
    C_{ij} = \Omega^2 \frac{Nd}{K{\cal N}} \sum_{ki'j'} \left\langle  h_{ii'}^{(k)}h_{jj'}^{(k)}\xi_{ii'}^{(k)}\xi_{jj'}^{(k)} \right\rangle,
\end{equation}
where averaging is performed over random numbers $\xi_{ij}^{(k)}$ and random placement of $\varkappa N$ bonds.

For non-diagonal elements $i\neq j$ the averaging of $\xi_{ii'}^{(k)}\xi_{jj'}^{(k)}$ is nonzero only if $i'=j$ and $j'=i$:
\begin{equation}
    C_{ij} = -\Omega^2 \frac{Nd}{K{\cal N}} \sum_{k} \left\langle  h_{ij}^{(k)}\right\rangle.
\end{equation}
After averaging over the random placement of bonds, we obtain $C_{ij}=-\Omega^2$ if $i$ and $j$ are neighbors and zero otherwise. For diagonal elements $i=j$ we obtain the usual sum rule $C_{ii} = -\sum_{j\neq i} C_{ij}$ because the sum rule $\sum_i A_{ik}=0$ is valid for any realization of the matrix $\hat{A}$.

The case $R=1$ coincides with a simple numerical model described in Section~\ref{sec:num}.

~\\
\section{Analytical form of the critical horizon}\label{append:Z(w)}

Using the small-argument expansion of $F_0(Z)$~(\ref{eq:asymp}), we obtain the conformal map~(\ref{eq:CH}) in the following form:
\begin{equation}
    \varkappa Z - a^2 \frac{Z^2}{\Omega^2} + \frac{Z^2\sqrt{-Z}}{4\pi\Omega^3} = \omega^2.   \label{eq:CH4}
\end{equation}
We solve Eq.~(\ref{eq:CH4}) using the method of simple iterations with the initial approximation $Z_0 = 0$ and recurrence relation
\begin{equation}
    \frac{1}{Z_{n+1}(\omega^2)} = \frac{\varkappa}{2\omega^2} + \frac{1}{\omega}\sqrt{f(\omega)+\frac{\sqrt{-Z_n(\omega^2)}}{4\pi\Omega^3}},
\end{equation}
obtained from Eq.~(\ref{eq:CH4}). The function $f(\omega)$ is defined in Eq.~(\ref{eq:f}). After each step, we keep only the leading terms of $\omega/\Omega$ and $\omega_c/\Omega$ taking into account that $\omega/\omega_c$ is arbitrary. After two iterations, this procedure converges to $Z(\omega^2)$ given in Eq.~(\ref{eq:Z(w)}).

\section{Non-Gaussian probability distribution}\label{sec:exp}

\begin{figure}[tb!]
    \includegraphics[scale=0.65]{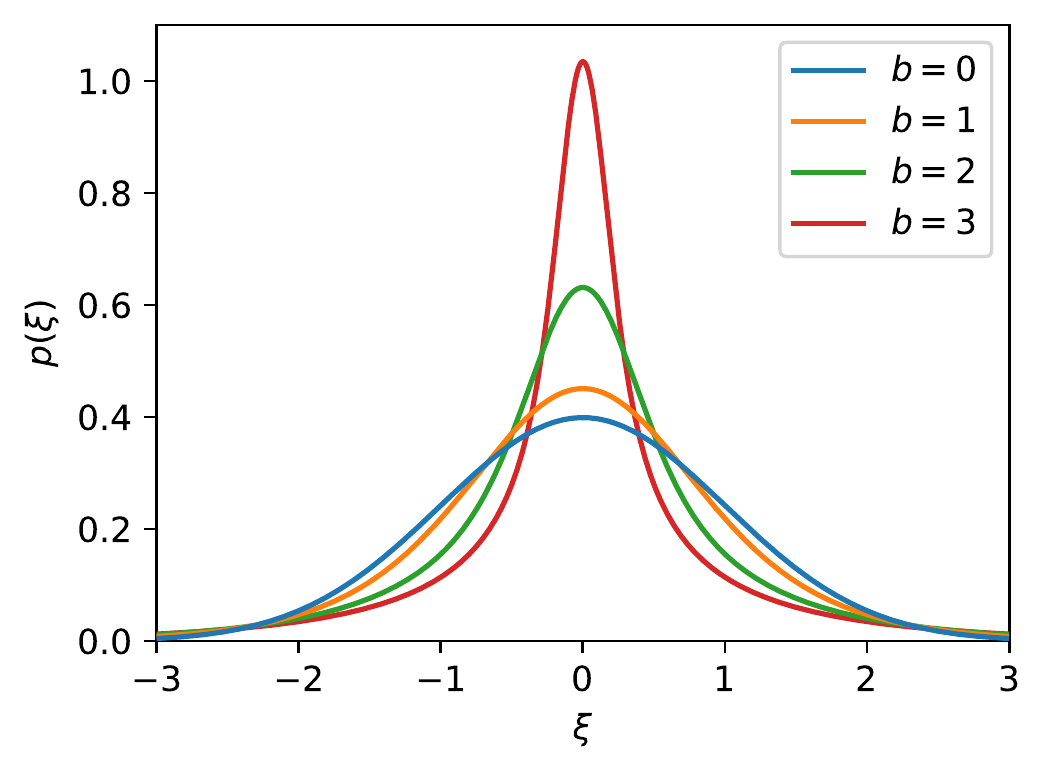}
    \caption{(Color online) Probability density function of the random variable $\xi$ defined by Eq.~(\ref{eq:rand_b}) for different values of the non-Gaussian parameter $b$.}
    \label{fig:exp}
\end{figure}

In order to simulate a wide range of different magnitudes of nonzero matrix elements, we multiply the standard Gaussian random number by log-uniform random variable. In this case the random variable has a form
\begin{equation}
    \xi = c\eta \exp{\zeta},  \label{eq:rand_b}
\end{equation}
where $\eta$ is the Gaussian random number with zero mean and unit variance and $\zeta$ is an independent random number uniformly distributed in the interval $(-b/2, b/2)$. The normalization constant $c = \sqrt{b/\sinh b}$ is chosen to provide the unit variance of $\xi$.

The resulting probability density function of the random variable $\xi$ has a form (see Fig.~\ref{fig:exp})
\begin{equation}
    p(\xi) = \frac{\operatorname{erf}\left(\frac{\xi e^{b/2}}{\sqrt{2b/\sinh b}}\right) - \operatorname{erf}\left(\frac{\xi e^{-b/2}}{\sqrt{2b/\sinh b}}\right)}{2b \xi}.
\end{equation}
For $b\ll 1$ it is close to the Gaussian distribution. However, with the increasing of $b$, the distribution of $\xi$ becomes closer to the reciprocal distribution $\sim\!1/|\xi|$.
The similarity to the Gaussian distribution can be quantitatively described as $\langle\xi^4\rangle/3\langle\xi^2\rangle^2$. For the Gaussian distribution, it is equal to 1. For the distribution $p(\xi)$, we obtain
\begin{equation}
    \frac{\langle\xi^4\rangle}{3\langle\xi^2\rangle^2} = \frac{b}{\tanh b},
\end{equation}
which differs significantly from 1 for $b\gg 1$.

%\bibliography{references}

\begin{thebibliography}{99}
    \bibitem{Allen 1993}
        P.\,B.\,Allen, and J.\,L.\,Feldman,
        Phys.\,Rev.\,B {\bf 48}, 12581 (1993).
    
    \bibitem{Allen 1999}
        B.\,Allen, J.\,L.\,Feldman, J.\,Fabian, and F.\,Wooten,
        Phil. Mag. B {\bf 79}, 1715 (1999).
    
    %%%%%%%%%% BOSON PEAK EXPERIMENTAL
    %%%%%%%%%%%%%%%%%%%%%%%%%%%%%%%%%%%%%%%%%
    \bibitem{Malinovsky 1986}
        V.\,K.\,Malinovsky, and A.\,P.\,Sokolov,
        %The nature of boson peak in Raman scattering in glasses
        Solid State Commun. {\bf 57}, 757 (1986).
    
    \bibitem{Kabeya 2016}
        M.\,Kabeya, T.\,Mori, Y.\,Fujii, A.\,Koreeda, B.\,W.\,Lee, J-H.\, Ko, and S.\,Kojima,
        %Boson peak dynamics of glassy glucose studied by integrated terahertz-band spectroscopy
        {Phys. Rev. B} {\bf 94}, 224204 (2016).
    
    \bibitem{Benassi 1996}
        P.\,Benassi, M.\,Krisch, C.\,Masciovecchio, V.\,Mazzacurati, G.\,Monaco, G.\,Ruocco, F.\,Sette, and R.\,Verbeni,
        %Evidence of High Frequency Propagating Modes in Vitreous Silica
        {Phys. Rev. Lett.} {\bf 77}, 3835 (1996).
    
    \bibitem{Wischnewski 1998}
        A.\,Wischnewski, U.\,Buchenau, A.\,J.\,Dianoux, W.\,A.\,Kamitakahara, and J.\,L.\,Zarestky,
        %Neutron scattering analysis of low-frequency modes in silica
        {Phil. Mag. B} {\bf 77}, 579 (1998).
    
    \bibitem{Matsuishi 1986}
        K.\,Matsuishi, S.\,Onari, and T.\,Arai,
        Jpn. J. Appl. Phys. {\bf 25}, 1144 (1986).
    
    \bibitem{Hutt 1989}
        K.\,W.\,Hutt, W.\,A.\,Phillips, and R.\,J.\,Butcher,
        J. Phys. Condens. Matter {\bf 1}, 4767 (1989).
    
    \bibitem{Ohsaka 1994}
        T.\,Ohsaka, and T.\,Ihara,
        Phys. Rev. B {\bf 50}, 9569 (1994).
    
    \bibitem{Zeller 1971}
        R.\,C. Zeller and R.\,O.\,Pohl,
        Phys. Rev. B {\bf 4}, 2029 (1971).
    
    \bibitem{Phillips 1981}
        W.\,A.\,Phillips, {\em Amorphous Solids: Low-Temperature Properties}
        (Springer, Berlin, 1981).
    
    \bibitem{White 1984}
        G.\,K.\ White, S.\,J.\,Collocott, and J.\,S.\,Cook,
        %Thermal expansion and heat capacity of vitreous B2O3
        Phys. Rev. B {\bf 29}, 4778 (1984).
    
    \bibitem{Kojima 2011}
        S.\,Kojima, Y.\,Matsuda, M.\,Kodama, H.\,Kawaji, and T.\,Atake,
        %Boson Peaks and Excess Heat Capacity of Lithium Borate Glasses
        {Chinese J. Phys.} {\bf 49}, 414 (2011).
    
    \bibitem{Steurer 2007}
        W.\,Steurer, A.\,Apfolter, M.\,Koch, W.\,E.\,Ernst, B.\,Holst, E.\,S\o{}nderg{\aa}rd, and J.\,R.\,Manson,
        %Observation of the Boson Peak at the Surface of Vitreous Silica //
        Phys. Rev. Lett. {\bf 99}, 035503 (2007).
    
    \bibitem{Steurer 2008-1}
        W.\,Steurer, A.\,Apfolter, M.\,Koch, W.\,E.\,Ernst, E.\,Sondergard, J.\,R.\,Manson, and B.\,Holst,
        Phys. Rev. B {\bf 78}, 045427 (2008).
    
    \bibitem{Steurer 2008-2}
        W.\,Steurer, A.\,Apfolter, M.\,Koch, W.\,E.\,Ernst, E.\,Sondergard, J.\,R.\,Manson, and B.\,Holst,
        Phys. Rev. Lett. {\bf 100}, 135504 (2008).

    \bibitem{Zhang 2017}
        L. Zhang, J. Zheng, Y. Wang, L. Zhang, Z. Jin, L. Hong, Y. Wang, and J. Zhang, Nat. Commun. {\bf 8}, 67 (2017).

    \bibitem{Wang 2018}
        Y. Wang, L. Hong, Y. Wang, W. Schirmacher, and J. Zhang, Phys. Rev. B {\bf 98}, 174207 (2018).
    %%%%%%%%%%%%%%%%%%%%%%%%%%%%%%%%%%%%%%%%%%%%%%%%%%%%%%%%55
    
    
    %%%%%%%%%%%%%%%%%%%%%%%%%%%%%%%%%%%%%%%%%
    
    \bibitem{Ruffle 2006}
        B.\,Ruffl\'e, G.\,Guimbreti\`ere, E.\,Courtens, R.\,Vacher, and G.\,Monaco,
        {Phys. Rev. Lett.} {\bf 96}, 045502 (2006).
    
    \bibitem{Ruffle 2008}
        B.\,Ruffl\'e, D.\,A.\,Parshin, E.\,Courtens, and R.\,Vacher,
        {Phys. Rev. Lett.} {\bf 100}, 015501 (2008).
    
    \bibitem{Shintani 2008}
        H.\,Shintani and H.\,Tanaka,
        {Nature Materials} {\bf 7}, 870 (2008)

    %%%%%%%%%% BOSON PEAK Theory
    %%%%%%%%%%%%%%%%%%%%%%%%%%%%%%%%%%%%%%%%%
    \bibitem{DeGiuli 2014}
        E.\,DeGiuli, A.\,Laversanne-Finot, G.\,Diring, E.\,Lernera, and M.\,Wyart,
        Soft Matter {\bf 10}, 5628 (2014).
    
    \bibitem{DeGiuli 2015}
        E.\,DeGiuli, E.\,Lerner, and M.\,Wyart,
        J. Chem. Phys. {\bf 142}, 164503 (2015).
    
    \bibitem{Wyart 2010}
        M.\,Wyart,
        EPL {\bf 89}, 64001 (2010).
    
    \bibitem{Wyart 2005}    
        M.\,Wyart, L.\,E.\,Silbert, S.\,R.\,Nagel, and T.\,A.\,Witten. Phys. Rev. E {\bf 72}, 051306 (2005).

    \bibitem{Schirmacher 2007}
        W.\,Schirmacher, G.\,Ruocco, and T.\,Scopigno,
        %Acoustic attenuation in glasses and its relation with the boson peak.
        Phys. Rev. Lett. {\bf 98}, 025501 (2007).
    
    \bibitem{Marruzzo 2013}
        A.\,Marruzzo, W.\,Schirmacher, A.\,Fratalocchi, and G.\,Ruocco, %Heterogeneous shear elasticity of glasses: The origin of the boson peak.
        Sci. Rep. {\bf 3}, 1407 (2013).
    
    \bibitem{Gurevich 2003}
        V.\,L.\,Gurevich, D.\,A.\,Parshin, and H.\,R.\,Schober,
        {Phys. Rev.} B {\bf 67}, 094203 (2003).
    
    \bibitem{Parshin 2007}
        D.\,A.\,Parshin, H.\,R.\,Schober, and V.\,L.\,Gurevich, Phys.\,Rev.\,B {\bf 76}, 064206 (2007).
    
    \bibitem{Karpov 1983}
        V.\,G.\,Karpov, M.\,I.\,Klinger, and F.\,N.\,Ignatev,
        Sov. Phys. JETP {\bf 57}, 439 (1983).
    
    \bibitem{Buchenau 1991}
        U.\,Buchenau, Yu.\,M.\,Galperin, V.\,L.\,Gurevich, and H.\,R.\,Schober,
        Phys. Rev. B {\bf 43}, 5039 (1991).
    
    \bibitem{Buchenau 1992}
        U.\,Buchenau, Yu.\,M.\,Galperin, V.\,L.\,Gurevich, D.\,A.\,Parshin, M.\,A.\,Ramos, and H.\,R.\,Schober,
        Phys. Rev. B {\bf 46}, 2798 (1992).


    \bibitem{Geotze 2000}
        W.\,G\"otze, and M.\,R.\,Mayr,
        Phys. Rev. E {\bf 61}, 587 (2000).        
    
    \bibitem{Taraskin 2001}
        S.\,N.\,Taraskin, Y.\,L.\,Loh, G.\,Natarajan, and S.\,R.\,Elliott,
        Phys.\,Rev.\,Lett. {\bf 86}, 1255 (2001).
    
    \bibitem{Chumakov 2011}
        A.\,I.\,Chumakov, G.\,Monaco, A.\,Monaco, W.\,A.\,Crichton, A.\,Bosak, R.\,R\"uffer, et al.,
        Phys. Rev. Lett. {\bf 106}, 225501 (2011).
    
    \bibitem{Tanaka 2008}
        H.\,Shintani,  and H.\,Tanaka,
        Nature Materials {\bf 7}, 870 (2008).
        %Universal link between the boson peak and transverse phonons in glass

    \bibitem{Milkus 2016}
        R.\,Milkus, and A.\,Zaccone,
        Phys.\,Rev.\,B {\bf 93}, 094204 (2016).  
        
    
    
    %%%%%%%%%%%%%%% Applications of RMT
    \bibitem{Speicher 2012}
        R.\,Speicher, and C.\,Vargas,
        %Free deterministic equivalents, rectangular random matrix models, and operator-valued free probability theory.
        {Random Matrices: Theory and Applications} {\bf 1}, 1150008 (2012).
    
    \bibitem{Prahofer 2000}
        M.\,Prahofer, and H.\,Spohn,
        %Universal Distributions for Growth Processes in 1+1 Dimensions and Random Matrices.
         {Phys. Rev. Lett.} {\bf 84}, 4882 (2000).
    
    \bibitem{Laloux 2000}
        L.\,Laloux, P.\,Cizeau, M.\,Potters, and J-P.\,Bouchaud,
        %Random matrix theory and financial correlations
        {Int. J. Theor. Appl. Finance} {\bf 3}, 391 (2000).
    
    \bibitem{Rajan 2006}
        K.\,Rajan, and L.\,F.\,Abbott,
        %Eigenvalue Spectra of Random Matrices for Neural Networks.
        {Phys. Rev. Lett.} {\bf 97}, 188104 (2006).
    
    \bibitem{Harnad 2011}
        J.\,Harnad,
        {\it Random Matrices, Random Processes and Integrable Systems} (Springer, New York, 2011).
    
    \bibitem{Tulino 2004}
        A.\,M.\,Tulino, and S.\,Verdu,
        %Random Matrix Theory and Wireless Communications.
        {Found. Trends Commun. Inf. Theory} {\bf 1}, 1 (2004).
    
    \bibitem{Wage 2015}
        K.\,E.\,Wage,
        %Application of random matrix theory to acoustic modeling and signal processing
        {J. Acoust. Soc. Am.} {\bf 138}, 1840 (2015).
    
    %\bibitem{Edelman 2014} Edelman A, Sutton B, Wang Y 2014
        %Random matrix theory, numerical computation and applications.
    %    {\it Proceedings of Symposia in Applied Mathematics} {\bf 72}
    
    \bibitem{Meyer 1997}
        H.\,Meyer, and J.\,C.\,Angles d'Auriac,
        %Random matrix theory and classical statistical mechanics: Spin models
        {Phys. Rev. E} {\bf 55}, 6608 (1997).
    
    \bibitem{Guhr 1998}
        T.\,Guhr, A.\,Mueller-Groeling, and H.\,A.\,Weidenmueller,
        %Random Matrix Theories in Quantum Physics: Common Concepts.
        {Phys. Rep.} {\bf 299}, 189 (1998).
    
    \bibitem{Olekhno 2018}
    N.\,A. Olekhno, Y.\,M.\,Beltukov. Phys. Rev. E {\bf 97}, 050101(R) (2018).
    
    %\bibitem{Jalan 2007} Jalan S, Bandyopadhyay J N 2007
        %Random matrix analysis of complex networks
    %    {\it Phys. Rev. E} {\bf 76} 046107
    %%%%%%%%%%%%%%%%%%%%%%%%
    
     
    
    \bibitem{Grigera 2002}
        T.\,S.\,Grigera, V.\,Martin-Mayor, G.\,Parisi, and P.\,Verrocchio,
        {J. Phys.: Cond. Matt.} {\bf 14}, 2167 (2002).
    
    \bibitem{Manning 2015}
        M.\,L.\,Manning, and A.\,J.\,Liu,
        {Europhysics Letters} {\bf 109}, 36002 (2015).
    
    \bibitem{Beltukov 2013}
        Y.\,M.\,Beltukov,  V.\,I.\,Kozub, and D.\,A.\,Parshin, Phys. Rev. B {\bf 87} 134203 (2013).

    \bibitem{Baggioli 2019}
        M.\,Baggioli, R.\,Milkus, and A.\,Zaccone,
        Phys. Rev. E {\bf 100}, 062131 (2019).            
    
    \bibitem{Beltukov 2015}
        Y.\,M.\,Beltukov,
        {JETP Letters} {\bf 101}, 345 (2015).
        
    \bibitem{Liu 2010}
        A.\,J.\,Liu, S.\,R.\,Nagel. Annu. Rev. Condens. Matter Phys. {\bf 1}, 347 (2010).
        
    \bibitem{OHern 2003}
        C.\,S.\,O'Hern, L.\,E.\,Silbert, A.\,J.\,Liu, and S.\,R.\,Nagel,
        {Phys. Rev. E} {\bf 68}, 011306 (2003).            
    
    \bibitem{Bhatia}
        R.\,Bhatia,
        \textit{Positive Definite Matrices} (Princeton University Press, Princeton, 2007).
    
    \bibitem{Beltukov 2016}    
        Y.\,M.\,Beltukov, D.\,A.\,Parshin. Boson peak in various random-matrix models. JETP letters {\bf 104}, 552 (2016).

    \bibitem{Burda 2004}
        Z.\,Burda, A.\,G\"orlich, A.\,Jarosz, and J.\,Jurkiewicz,
        {Physica A} {\bf 343}, 295 (2004).        
    
    \bibitem{Maxwell 1865}
        J.\,C.\,Maxwell, Philosophical Magazine {\bf 27}, 294 (1865).   

    \bibitem{Stillinger 1985}
        F.\,H.\,Stillinger and T. A. Weber, Phys. Rev. B {\bf 31}, 5262 (1985).        

    \bibitem{Beltukov 2011} 
        Y.M. Beltukov and D.A. Parshin, Phys. Solid State {\bf 53}, 142 (2011).              

        
    \bibitem{Zucker 2011}
        I.\,J.\,Zucker,
        %70+ Years of the Watson Integrals,
        J. Stat. Phys. {\bf 145}, 591 (2011).

   
    \bibitem{Burda 2006}
        Z.\,Burda, A.\,G\"orlich, J.\,Jurkiewicz, B. Wac\l{}aw. Eur. Phys. J. B {\bf 49}, 319 (2006).
        
    
    \bibitem{SM}
        See Supplemental Materials (\href{https://arxiv.org/src/2007.12288/anc/vdos.gif}{vdos.gif} in the Ancillary files of the arXiv submission) for the animated plot of the VDOS.         
        
    \bibitem{Beltukov 2018}
        Y.\,M\,Beltukov. AIP Conf Proc {\bf 1978}, 030021 (2018).

    \bibitem{Beltukov 2016 PRE} Y. M. Beltukov, C. Fusco, D. A. Parshin, A. Tanguy. Phys. Rev. E {\bf 93} 023006 (2016).

        
    \bibitem{KPM 2006}
        A.\,Wei{\ss}e, G.\,Wellein, A.\,Alvermann, and H.\,Fehske,
        %The kernel polynomial method //
        {Rev. Mod. Phys.} {\bf 78}, 275 (2006).  
        
    \bibitem{Rainone 2020}
        C. Rainone,  E. Bouchbinder, E. Lerner. PNAS {\bf 117}, 5228 (2020).        

    \bibitem{Vitelli-2010}
        V.\,Vitelli, N.\,Xu, M.\,Wyart, A.\,J.\,Liu, and S.\,R.\,Nagel,
        %Heat transport in model jammed solids //
        Phys. Rev. E {\bf 81}, 021301 (2010).  
        
    % \bibitem{Scopigno 2006}
    %     T. Scopigno, J. B. Suck, R. Angelini, F. Albergamo,  G. Ruocco, Phys. Rev. Lett. {\bf 96}, 135501 (2006).        
    \bibitem{Narayan 2020}
        O. Narayan, H. Mathur, arXiv:2006.16497.


    \bibitem{Sette-1998}
        F. Sette, M. H. Krisch, C. Masciovecchio, G. Ruocco, and G. Monaco,
        Science {\bf 280}, 1550 (1998).
    
    \bibitem{Ruocco-2001}
        G.\,Ruocco, and F.\,Sette,
        J. Phys.: Cond. Mat. {\bf 13}, 9141 (2001).
    
    \bibitem{Christie-2007}
        J.\,K.\,Christie, S.\,N.\,Taraskin, and S.\,R.\,Elliott,
        J. Non-Cryst. Solids {\bf 353}, 2272 (2007).

   
    \bibitem{Monaco 2009a}
        G. Monaco, V. M. Giordano. PNAS {\bf 106}, 3659 (2009).

    \bibitem{Monaco 2009}
        G.\,Monaco, and S.\,Mossa, PNAS {\bf 106}, 16907 (2009).        

    % \bibitem{Schirmacher 1998}
    %     W.\,Schirmacher, G.\,Diezemann, and C.\,Ganter, Phys.\,Rev.\,Lett., {\bf 81}, 136 (1998).
    
    \bibitem{Mizuno 2017}
        H.\,Mizuno, H.\,Shibab, and A.\,Ikeda,
       %Continuum limit of the vibrational properties of amorphous solids.
       Proc. Natl. Acad. Sci. USA {\bf 114}, E9767 (2017).
    
    \bibitem{Mizuno 2018}
       H.\,Mizuno, and A.\,Ikeda, Phys. Rev. E {\bf 98}, 062612 (2018).
    
    % \bibitem{Klinger 2001}
    %     M.\,I.\,Klinger, and A.\,M.\,Kosevichb,
    %     Physics Letters A {\bf 280}, 365 (2001).
    

    % \bibitem{Conyuh 2017}
    %     D.\,A.\,Conyuh, Y.\,M.\,Beltukov, and D.\,A.\,Parshin, J. Phys. Conf. Ser. {\bf 929}, 012031 (2017).

    \bibitem{Conyuh 2020}
        D.\,A.\,Conyuh, Y.\,M.\,Beltukov. Phys. Solid State {\bf 62}, 689 (2020).        
 
        
    % \bibitem{Conyuh 2019}    
    %     D.\,A.\,Conyuh, Y.\,M.\,Beltukov,D.\,A.\,Parshin. J. Phys.: Conf. Ser. {\bf 1400}, 044026 (2019).  
    

    \bibitem{Lerner 2016}
        E. Lerner, G. D\"uring and E. Bouchbinder, Phys. Rev. Lett. {\bf 117}, 035501 (2016).

    %\bibitem{Mizuno 2017 }

    \bibitem{Shimada 2018}
        M. Shimada, H. Mizuno and A. Ikeda, Phys. Rev. E,
        2018, 97, 022609 (2018).

    \bibitem{Lerner 2017}
        E. Lerner and E. Bouchbinder, Phys. Rev. E {\bf 96}, 020104(R) (2017).

    \bibitem{Kapteijns 2018}
        G. Kapteijns, E. Bouchbinder and E. Lerner, Phys. Rev. Lett. {\bf 121}, 055501 (2018).

    \bibitem{Wang 2019}
        L. Wang, A. Ninarello, P. Guan, L. Berthier, G. Szamel and
        E. Flenner, Nat. Commun. {\bf 10}, 26 (2019).    
      
    \bibitem{Shimada 2020}
        M. Shimada, H. Mizuno, A. Ikeda. Soft Matter {\bf 16}, 7279 (2020).


    \bibitem{Charbonneau 2016} 
    P. Charbonneau, E. I. Corwin, G. Parisi, A. Poncet, and F. Zamponi, Phys. Rev. Lett. {\bf 117}, 045503 (2016). 
    
    \bibitem{Ji 2019} 
    W. Ji, M. Popovic, T. W. J. de Geus, E. Lerner, and M. Wyart, Phys. Rev. E {\bf 99}, 023003 (2019).
    
    % \bibitem{Marchenko 1967}
    %     V.\,A.\,Marchenko, and L.\,A.\,Pastur,
    %     {Mat. Sbornik} {\bf 72}, 507 (1967).
    
    %%%%%%%%%%%%%%%%%%%%%%%%%%%%%%%%%%%55
    

    
    % \bibitem{Marchenko 1967}
    %     V. A. Marchenko, and L. A. Pastur, Mat. Sbornik {\bf 72}, 507 (1967).

        



        

        
        

    %\bibitem{Gurevich-1993}  Gurevich V L , Parshin D A, Pelous J and Schober H R 1993 {\it Phys. Rev.} B {\bf 48} 16318
    
    %\bibitem{Parshin-2001} Parshin D A and Laermans C 2001 {\it Phys. Rev.} B {\bf 63} 132203
    
    %\bibitem{Maradudin 1971} Maradudin A A, Montroll E W, Weiss G H and Ipatova I P 1971 {\it Theory of Lattice Dynamics in the Harmonic Approximation} (New York: Academic Press)
    

    
    %\bibitem{Beltukov 2013} Beltukov Y M, Kozub V I and Parshin D A 2013 {\it Phys. Rev.} B {\bf 87} 134203
    
    %\bibitem{Burda 2006} Burda Z, Gorlich A, Jurkiewicz J and Waclaw B 2006 {\it Eur. Phys. J.} B {\bf 49} 319
    
    %\bibitem{Wyart 2010} Vitelli V, Xu N, Wyart M, Liu A J, and Nagel S R 2010 {\it Phys. Rev.} E {\bf 81} 021301
    
    %\bibitem{Kojima 1999} Kojima S and Kodama M 1999 {\it Physica B: Cond. Mat.} {\bf 263} 336
    
    %\bibitem{Beltukov 2016} Beltukov Y M and Parshin D A 2016 {\it JETP Letters} {\bf 104} 552
    
    %\bibitem{Beltukov 2016 PRE} Beltukov Y M, Fusco C, Parshin D A and Tanguy A 2016 {\it Phys. Rev.} E {\bf 93} 023006
    
    %\bibitem{Weisse 2006} Weisse A, Wellein G, Alvermann A and Fehske H 2006 {\it Rev. Mod. Phys.} {\bf 78} 275
    
    
\end{thebibliography}

\end{document}